\documentclass[preprint,10pt,3p,onecolumn]{elsarticle} 

\usepackage{xcolor}
\usepackage{amssymb}
\usepackage{amsmath}
\usepackage{url}
\usepackage{rotating}
\usepackage{mathtools, cuted}
\usepackage{float}
\usepackage{textcomp}
\usepackage{listings}

\usepackage{algorithm}
\usepackage{algpseudocode}

\usepackage[normalem]{ulem}

\journal{Astronomy \& Computing}

\newcommand{\cosmokdtree}[1]{\textsc{cosmokdtree}}
\newcommand{\scipy}[1]{\textsc{scipy}}
\newcommand{\coretran}[1]{\textsc{coretran}}
\newcommand{\torchkdtree}[1]{\textsc{torch\_kdtree}}
\usepackage{esint}

\begin{document}

\begin{frontmatter}

\title{\cosmokdtree{}: a flexible OpenMP-parallelized $k$-d tree for computational astrophysics applications}

\author[inst1]{\'{O}scar Monllor-Berbegal} \corref{cor1} \ead{oscar.monllor@uv.es} 
\author[inst3,inst4,inst1]{David Vall\'{e}s-P\'{e}rez}
\author[inst1,inst2]{Susana Planelles}
\author[inst1,inst2]{Vicent Quilis}

\address[inst1]{Departament d'Astronomia i Astrof\'{\i}sica, Universitat de
  Val\`encia, E-46100 Burjassot (Val\`encia), Spain}
\address[inst3]{Dipartimento di Fisica e Astronomia, Università di Bologna, Via Piero Gobetti 93/2, IT-40129 Bologna, Italy}
\address[inst4]{Istituto di Radioastronomia -- Istituto Nazionale di Astrofisica (IRA-INAF), Via Piero Gobetti 101, IT-40129 Bologna, Italy}
\address[inst2]{Observatori Astron\`omic, Universitat de Val\`encia, E-46980
  Paterna (Val\`encia), Spain}

\begin{abstract}
Modern numerical astrophysics applications present a common demand for efficient spatial querying techniques, such as neighbor searches or density estimations over billions of resolution elements. To address the growing need for fast and versatile tools tailored to these tasks, we present \cosmokdtree{}, a fast and flexible multi-purpose $k$-d tree implementation in \texttt{Fortran}, parallelized with \texttt{OpenMP} directives. Our library supports arbitrary dimensionality and spatial distributions, and is designed for efficient tree construction and fast query performance for moderate-scale applications (performance tested up to $\sim 10^9$ points). \texttt{Python} bindings coupled to the module are also provided. All these ingredients yield  a $k$-d tree package suitable for analysis purposes.

We benchmark \cosmokdtree{} across a variety of scenarios, including different point distributions, dimensionality, and parallel scaling. We also compare its performance against the widely used \scipy{} implementation and other efficient alternatives such as \coretran{}'s and a GPU-based version, showing that \cosmokdtree{} consistently achieves lower construction and query times than CPU alternatives while keeping a reasonable memory usage. Our tree-building phase implementation, executed on a mid-range workstation-class CPU, approaches the performance of GPU-based implementations when run on high-end consumer graphic cards, although data-center GPUs, out of the scope for our comparison, could still deliver substantially higher performance and a broader conclusion cannot be extracted. We further present some applications tackling common problems in astrophysics, namely, the friends-of-friends clustering algorithm and the particle-to-mesh assignment process.
The code is publicly released and intended to serve as a flexible multi-purpose tool for computational applications in a wide range of scenarios, particularly in astrophysics.
\end{abstract}


\end{frontmatter}

\section{Introduction}
\label{s:intro}

Spatial querying operations, such as the identification of the closest neighbors to a point (a \textit{nearest neighbors search}) or finding all elements intersecting a certain volume (a \textit{range search}), are ubiquitous in numerical computations generating or analyzing scientific data. These applications are fundamental in a wide range of scientific fields ranging from astrophysics \citep{knebe2011haloes,saftly2014hierarchical} to molecular dynamics \citep{chen2017using}, image rendering \citep{hapala2011kd}, medical physics \citep{quiroga2004unsupervised,andreopoulos2007clustering}, or machine learning \citep{pelleg1999accelerating, chatterjee2025learning}, among many others. 
Since datasets are increasingly growing in size and complexity (either being simulation outputs, object catalogues, etc.), the use of efficient methods that significantly reduce computational costs becomes mandatory. A prototypical example of such needs could be numerical cosmology, where modern simulations often involve unstructured data representations (e.g. particles in $N$-body or SPH simulations, moving-meshes, etc.) with a large quantity of numerical elements (beyond $10^{9}$) to solve the equations governing the evolution of the physical processes involved. In this sense, extracting information from such sophisticated datasets (e.g., using structure identification algorithms such as halo finders \citep{knebe2011haloes} or void finders \citep{colberg2008aspen}) poses a significant computational challenge.

A naive brute force approach can be easily applied to carry out the neighbor and range searches; yet, if the number of points is increased, it can rapidly become prohibitive in terms of wall time ($\mathcal{O}[N^2]$ complexity). In this direction, more advanced techniques exist for accelerating these calculations. The Hilbert curve \citep{sagan1994hilbert}, for example, is a space-filling curve that maps a multi-dimensional space onto a one-dimensional one, preserving locality. Points close together in the original space will also be close (clustered) when mapped to the Hilbert curve's one-dimensional representation and, thus, can be applied to find the neighbors of a point. On the other hand, space-partitioning tree data structures, used to spatially index objects by recursively subdividing the input domain, are also a proper choice to reduce computational costs. Among these, one can choose from several options: \textit{k}-d trees \citep{bentley1975kdtree}, Octrees \citep{meagher1980octree}, $R^{*}$-trees \citep{beckmann1990r}, \textit{X}-trees \citep{berchtold1996x}, just to mention a few. The best choice depends on the application. Octrees are particularly well-suited for indexing three-dimensional point data due to their hierarchical decomposition of space into uniform cubic regions. In contrast, $R^*$-trees are more effective for indexing spatially extended objects (e.g. rectangles or polygons). The $X$-tree, being an extension of the $R^*$-tree, offers improved performance in high-dimensional spaces by minimizing overlap and excessive node splitting. Meanwhile, $k$-d trees are highly efficient for nearest-neighbor and range searches in low-dimensional point datasets due to their axis-aligned, adaptive binary partitioning strategy.

First introduced by \citet{bentley1975kdtree}, the $k$-d tree is a binary tree in which every node is a point in a $k$-dimensional space that divides it, across one axis, into two parts: the left and the right subtrees. The splitting technique, i.e. the choice of the point and axis dividing data in two halves at each level of the tree hierarchy, should ensure balance, guaranteeing fast tree traversals. The \mbox{$k$-d} tree construction time complexity for $N$ input points is $\mathcal{O}(N\log N)$. After that, traversing the tree to find a given element becomes $\mathcal{O}(\log N)$, compared to the naive $\mathcal{O}(N)$ complexity. More details on $k$-d tree construction and queries are provided in Sec. \ref{s:algorithm}.

Many efficient $k$-d tree implementations already exist in the literature for various programming languages: \texttt{C++} \citep{arya1998optimal, muja2014scalable}, \texttt{Fortran} \citep{coretran} or \texttt{Python} \citep{pedregosa2011scikit,virtanen2020scipy}. Besides, several versions for GPU-based parallel construction and querying have also been presented in Refs. \citep{hu2015massively, KdTreeGPU, wald2022gpu}. In this scenario, with a variety of efficient $k$-d tree packages already released, we present \cosmokdtree{}, a new implementation of the algorithm aimed to maximize tree construction and query efficiency while keeping flexibility and computational costs low without resorting to GPUs, as our goal is not massive parallel efficiency, which often comes at the expense of flexibility.   This new version is written in \texttt{Fortran} and parallelized by means of \texttt{OpenMP} (\texttt{OMP}) directives, being able to build the tree for $\sim 10^9$ points in 3-dimensional space in less than a minute. It is furthermore a highly flexible tool, as the user can specify the integer size, floating point arithmetics precision, periodic boundary conditions or lack thereof, or dimensionality of the input space at compilation time, depending on the intended application, without a significant performance decrease when varying these parameters. Besides the \texttt{Fortran} module, a \texttt{Python} version is also provided without significant overhead added to the building and querying phases. This fact increases its usability in moderate-scale data-analysis scenarios.

Though first conceived for cosmological simulation output analysis, \cosmokdtree{} could be widely applied in all areas encompassed by computational physics. It could help analyze \citep{saftly2014hierarchical, valles2024vortex, monllor2025avism} or accelerate \citep{chen2017using, frontiere2025cosmological}  $N$-body simulations. Regarding the identification of groups or structures, it could help optimize clustering algorithms, \citep{springel2001populating,gao2008support,behroozi2012rockstar}, catalog cross-matching \citep{robitaille2013astropy} or correlation function calculations \citep{zhao2023fast}. Furthermore, machine learning algorithms could also benefit from using a $k$-d tree, as it can facilitate classification, regression, or anomaly detection utilizing its fast neighbor and range searches (e.g., see \citet{pelleg1999accelerating, chatterjee2025learning}).

The rest of the manuscript is structured as follows: in Sec. \ref{s:algorithm} we give algorithmic details about our particular implementation, both in tree construction and queries. In Sec. \ref{s:benchmarks} we asses \cosmokdtree{}'s performance, providing benchmarks of performance against the number of points, their spatial distribution, dimensionality and CPU architecture, and comparing to other widely used implementations. In Sec. \ref{s:application}, some applications of our code, especially in the field of Cosmology, are described. Finally, in Sec. \ref{s:summary}, we provide a brief summary of the algorithm and its implications. In \ref{s:appendix.1} a study on the choice of the best leaf size is provided,  whereas in \ref{s:appendix.2}, we provide a short review on tree construction techniques.

\begin{table*}[h!]
\centering
\caption{Index of symbols and brief description}
\label{tab:symbols}
\begin{tabular}{c|l}
\hline
\textbf{Symbol} & \textbf{Description} \\
 \hline  \hline 
$N$ & Number of input points \\
$D$ & Dimensionality of the input space \\
$\mathcal{S}$ & Input array storing $N$ points in a $D$-dimensional space \\
$\ell$ & Tree level of a given node\\
$N_\text{leaf}$ & Tree leaf size \\
$\mathbf{n}$ & Tree nodes saving the splitting data (splitting axis, point and child bounding boxes) \\
$\mathbf{n}^\text{L}, \mathbf{n}^\text{R}$ & Left and right children of $\mathbf{n}$\\
$\mathbf{n}_0$ & Root node from which all others are born \\
$P_\perp$ & Node's point splitting data in two parts at a given level \\
$i_\perp$ & Node's splitting axis \\
$x_\text{max}^{i}$& Node's bounding box upper bound across the $i$-th axis \\
$x_\text{min}^i$& Node's bounding box lower bound across the $i$-th axis \\
$\mathcal{S}_\ell$ & A subset of $\mathcal{S}$ at level $\ell$ resulting from the binary splitting logic ($\mathcal{S}_0 =\mathcal{S}$) \\
$\mathcal{S}_\ell^\text{L}, \mathcal{S}_\ell^\text{R}$ & Binary divisions of $\mathcal{S}_\ell$ that become $\mathcal{S}_{\ell+1}$ at the next level ($\ell+1$)\\
\hline
$N_\text{CPU}$ & Number of available threads to execute \texttt{OMP} jobs concurrently \\
$\ell_\text{max}$ & Maximum tree level to allow parallelization \\
\hline
$P$ & Query point \\
$R$ & Radius of the $(D-1)$-sphere (ball) used to query the tree \\
$k$ & Number of nearest neighbors to search \\
\end{tabular}
\end{table*}

\section{Algorithm}
\label{s:algorithm}

In this section, we introduce the $k$-d tree algorithm and we describe our particular implementation, discussing the particular techniques applied for accelerating the tree construction and queries. The code is fully written in \texttt{Fortran}, while parallelization steps are carried out using \texttt{OpenMP} instructions. For the sake of clarity, Table \ref{tab:symbols} provides a summary of the notation used throughout the whole manuscript for all variables involved in the $k$-d tree construction and queries.

\subsection{Tree construction}

\begin{figure*}[h!]
\centering 
\includegraphics[width=0.8\linewidth]{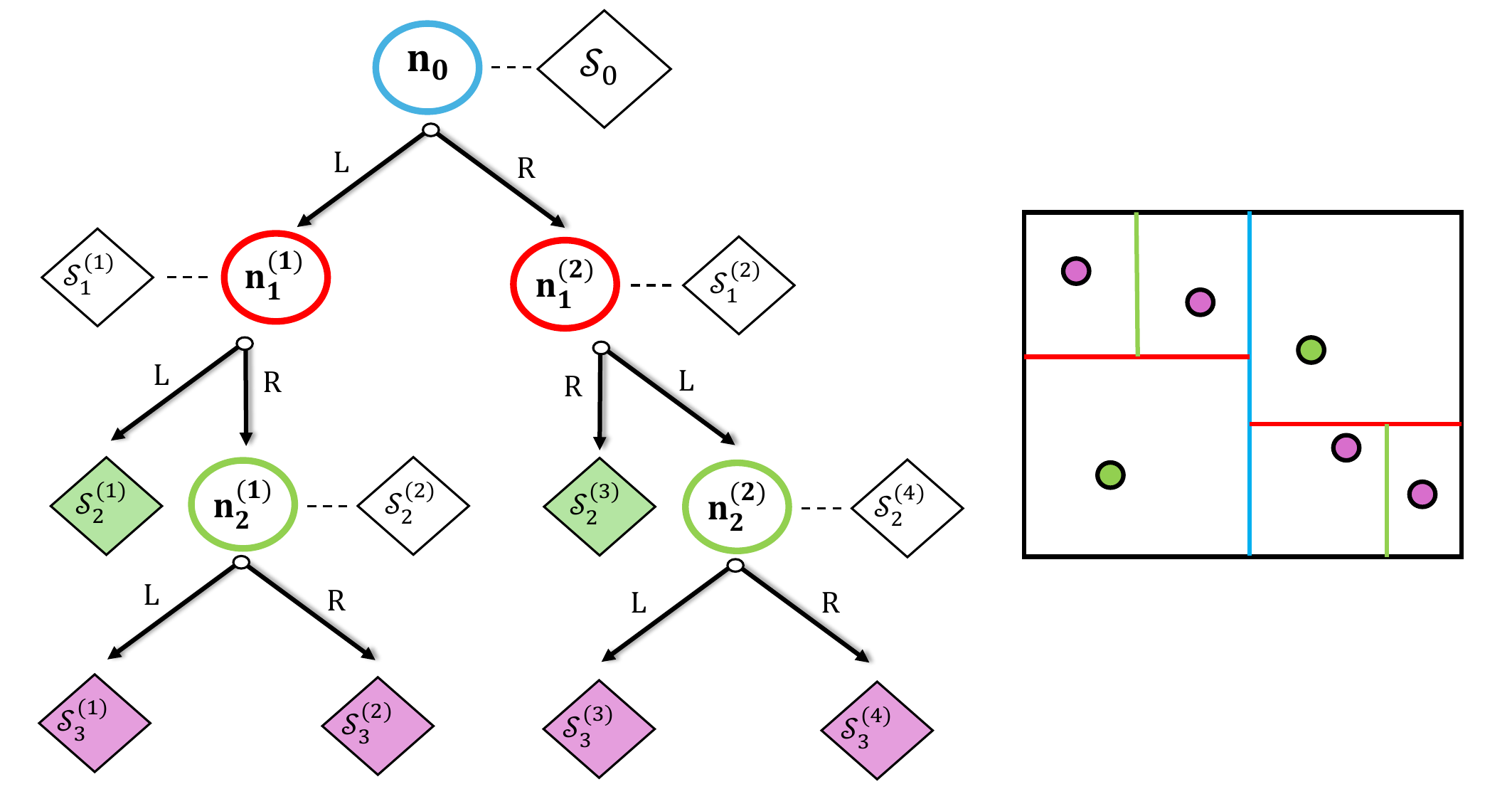}
\caption{\textit{Left:} diagram displaying a $k$-d tree for a 2D case with $N = 6$ points. Diamonds represent the sets and subsets that are subsequently split by the hyperplanes (in this case, straight lines) saved in the different tree nodes, depicted with colored circumferences. Leaf nodes are represented by colored diamonds. In this sense, colors indicate those elements that belong to the tree structure. Lower indices display tree level ($\ell$), while superscripts display node and subset unique identifiers within a given level. \textit{Right:} 2D sketch of the $N = 6$ input set of points ($S_0$) provided to the $k$-d tree. Lines (color-coded as in the \textit{left} figure) divide the data in the different levels of the tree. For higher dimensionalities, lines would be hyperplanes. Non-leaf nodes only save the splitting information, while leaf nodes are formed by the points with the regions delimited by the splitting lines. Again, colors used to draw the points correspond with the ones used in the \textit{left} diagram and indicate the level at which they are saved.}
\label{fig:kdtree_diagram}
\end{figure*}

Given a set $\mathcal{S}$ of $N$ points in a $D$-dimensional metric space $X$, we want to arrange all elements in a $k$-d tree \citep{bentley1975kdtree} structure, suitable for obtaining fast nearest neighbor and range queries. In our particular implementation, we assume $X = \mathbb{R}^D$ (real $D$-dimensional space) with distances measured according to an $L^p$ Minkowski norm,

\begin{equation}
\label{eq:minkowski}
\text{dist}(\mathbf{x}, \mathbf{y}) = \left( \sum_{i=1}^{D} |x_i - y_i|^p \right)^{\frac{1}{p}} \; \;  \text{for} \; \; \mathbf{x}, \mathbf{y} \in \mathbb{R}^D,
\end{equation}

where, by default, $p = 2$, although it can be easily modified. Furthermore, periodic boundary conditions are also implemented by replacing this metric with the one inside of a $D$-toroid.

With these considerations, below we provide our recursive splitting logic employed to produce a $k$-d tree structure from the input set of points ($\mathcal{S} \equiv \mathcal{S}_0$). To divide the input dataset and the subsequent subsets, we utilize the sliding-midpoint splitting method as described by \cite{maneewongvatana1999analysis} and similar to \citep{virtanen2020scipy}.  So as to provide a more visual understanding of the $k$-d structure and clarify our notation, Fig. \ref{fig:kdtree_diagram} displays a full tree hierarchy for a 2D case with $N = 6$ input points inside a rectangle. The algorithm proceeds as follows:

\begin{enumerate}
    \item In the first step ($\ell = 0$), the splitting axis ($i_\perp$) is chosen as the axis with the largest bounding box side, irrespective of the point distribution. The splitting point ($P_\perp$) does not need to belong to $\mathcal{S}_0$ and is chosen as the bounding box midpoint across axis $i_\perp$. If all points lie on one side of the resulting splitting hyperplane, it slides until it encounters the first point (hence the name of the method). In that case, the point stopping the sliding now belongs to the side that was empty when using only the midpoint logic, avoiding trivial splits that would result in an unnecessarily deeper tree.
    
    \item After setting the splitting hyperplane, $\mathcal{S}_0$ is divided into the left ($\mathcal{S}^L_0$) and right halves ($\mathcal{S}^R_0$). The splitting point and axis as well as the bounding box ($x_\text{min}^1,x_\text{max}^1, x_\text{min}^2,x_\text{max}^2,\dots,x_\text{min}^D,x_\text{max}^D$) become the initial node ($\mathbf{n}_0$) that is connected to the left and right child nodes that will further split both halves ($\mathcal{S}_1 = \mathcal{S}^L_0, \mathcal{S}^R_0$) at $\ell = 1$.
    
    \item At any further level ($\ell>0$), the same logic is recursively applied. In general, $\mathcal{S}_\ell$ at level $\ell$ is divided into $\mathcal{S}^L_\ell$ and $\mathcal{S}^R_\ell$ and a new node ($\mathbf{n}$), saving the subset and splitting information is created, connected to its left ($\mathbf{n}^\text{L})$ and right children ($\mathbf{n}^\text{R})$ at $\ell+1$, which further divide the left and right subsets of points\footnote{Note that, at each $\ell$, there can be as many $\mathcal{S}_\ell$ as $2^\ell$.}. By definition, these subsets fulfill the following relation:
    \begin{equation}
    \label{eq:left_right}
    \mathcal{S}_\ell = \operatorname{Concat}_{i_\perp}(\mathcal{S}_\ell^\text{L}, \mathcal{S}_\ell^\text{R})\, ,
    \end{equation}
    where $\operatorname{Concat}_{i_\perp}$ is the concatenation operator across axis ${i_\perp}$.
    \item The recursion stops when the cardinality of $\mathcal{S}_\ell$ is below a given quantity called \textit{leaf size} ($N_\text{leaf}$). In this case, the corresponding node saves all points in $\mathcal{S}_\ell$ and it no longer splits the data. 
\end{enumerate}

A $k$-d tree construction routine must ensure that the resulting tree is balanced (tree depth is limited) and complete (Eq. \eqref{eq:left_right} is always fulfilled). The sliding-midpoint (\cosmokdtree{}'s approach) explicitly sacrifices strict tree structural balance in exchange for better geometrical balance. This means that the maximum depth cannot be delimited, but the approach maximizes node efficiency and, ultimately, query performance. In contrast, the standard (perfectly balanced by definition) maximum-variance technique can yield an inefficient node structure if the point distribution is clustered along a given direction (e.g., see \citep{maneewongvatana1999analysis}).

An important feature affecting building speed is the \textit{leaf size} ($N_\text{leaf}$). It represents the maximum number of points to stop the splitting process and, thus, tree construction in that particular branch of the tree. If this condition is satisfied by a node, the tree construction routine simply saves the indices and positions of all points inside the node. When a query traverses the tree and finds a leaf, it applies brute force on the points saved in it. Hence, $N_\text{leaf}$ has to be carefully chosen to accelerate tree construction without affecting query performance. A large value of $N_\text{leaf}$ provides faster building times but makes queries slower due to the abuse of brute force. On the contrary, setting $N_\text{leaf}$ too low (for instance, as unity, as in \coretran{} \cite{coretran}) not only slows down tree construction, but can also downgrade query performance due to excessive subtree visits. After thorough testing on different numbers of input points and dimensions, we realized that the optimal $N_\text{leaf}$ depends non-trivially on these variables. In general, a leaf size in the range $N_\text{leaf} = 10-20$ yields significantly smaller tree construction times, while making queries slightly faster or, at least, equally efficient. Because of this reason, we will consider $16$ our default value. Larger values of $N_\text{leaf}$ can yield better or worse query performance depending on $N$ and $D$, and care should be taken when modifying the default value, as queries can get exponentially slower when raising $N_\text{leaf}$. For more details on this matter, we refer the reader to \ref{s:appendix.1}, where we provide a comprehensive study on how \cosmokdtree{}'s leaf size configuration affects its tree construction and query performance for varying $N$ and $D$.

In Algorithm \ref{alg:parallel}, we display pseudocode for our main tree construction routine with all the ingredients detailed above. A comprehensive description of the tree parallelization part, which is key for obtaining fast building times, is provided in the next section. It is also important to discuss the strengths and caveats of our methodology. As such, in \ref{s:appendix.2} an overview of the current state-of-the-art techniques applied to efficiently build $k$-d trees, both on CPU and GPU, is provided.

\vspace{0.3 cm}
\begin{algorithm}[!h]
\caption{$k$-d tree recursive construction routine.}
\label{alg:parallel}
\begin{algorithmic}[1]
\Procedure{BuildTree}{$\mathcal{S}_\ell,\ell,\ell_\text{max},N_\text{leaf}$}
\If{$\texttt{size}(\mathcal{S}_\ell) \leq N_\text{leaf}$}
    \State $\mathbf{n} \leftarrow \mathcal{S}_\ell$ (save all leaf points)
    \State \textbf{return} 
\EndIf
\State 
\State Calculate largest bounding box side ($i_\perp$) and its midpoint $P_\perp$
\State
\State Check whether the splitting plane should slide 
\State 
\State Save node splitting point, axis and child bounding boxes
\State
\If{$d < \ell_\text{max}$}
    \State \textbf{parallel} (2 threads)
    \State \quad \textbf{single}
    \State \quad\quad \textbf{task}
    \State \quad \quad \quad $\mathbf{n}(\mathbf{n}^\text{L}) \leftarrow$\Call{BuildTree}{$\mathcal{S}_\ell^\text{L}$, $\ell+1$, $\dots$}
    \State \quad\quad \textbf{end task}
    \State \quad\quad \textbf{task}
    \State \quad\quad \quad $\mathbf{n}(\mathbf{n}^\text{R}) \leftarrow$ \Call{BuildTree}{$\mathcal{S}_\ell^\text{R}$, $\ell+1$, $\dots$}
    \State \quad\quad \textbf{end task}
    \State \quad \textbf{end single}
    \State \textbf{end parallel}
\Else
    \State $\mathbf{n}(\mathbf{n}^\text{L}) \leftarrow$ \Call{BuildTree}{$\mathcal{S}_\ell^\text{L}$, $\ell+1$, $\dots$}
    \State $\mathbf{n}(\mathbf{n}^\text{R}) \leftarrow$ \Call{BuildTree}{$\mathcal{S}_\ell^\text{R}$, $\ell+1$, $\dots$}
\EndIf
\EndProcedure
\end{algorithmic}
\end{algorithm}

\subsection{Parallelization}
\label{s:parallelization}

Due to the intrinsic top-down sequential nature of the algorithm, parallelizing the tree construction part is inherently complicated. This is why many of the publicly available implementations build the tree serially (e.g. \citep{coretran}, \citep{virtanen2020scipy}). Nevertheless, by leveraging the \texttt{OMP task} construct and enabling \textit{nested parallelism}, a significant (although not ideal) speedup can be obtained. \cosmokdtree{}'s parallelization is fully carried out harnessing \texttt{OMP} directives (\texttt{parallel}, \texttt{single} and \texttt{task}) on a level-by-level basis, following the top-down construction approach. While this technique limits parallel scalability (see \ref{s:appendix.2}), it keeps the point partitioning scheme logic simple, efficient, and accessible\footnote{By accessible we mean that queries can access the $k$-d tree structure without performing extra operations.} by query routines.

\begin{figure}[h!]
\centering 
\includegraphics[width=0.7\linewidth]{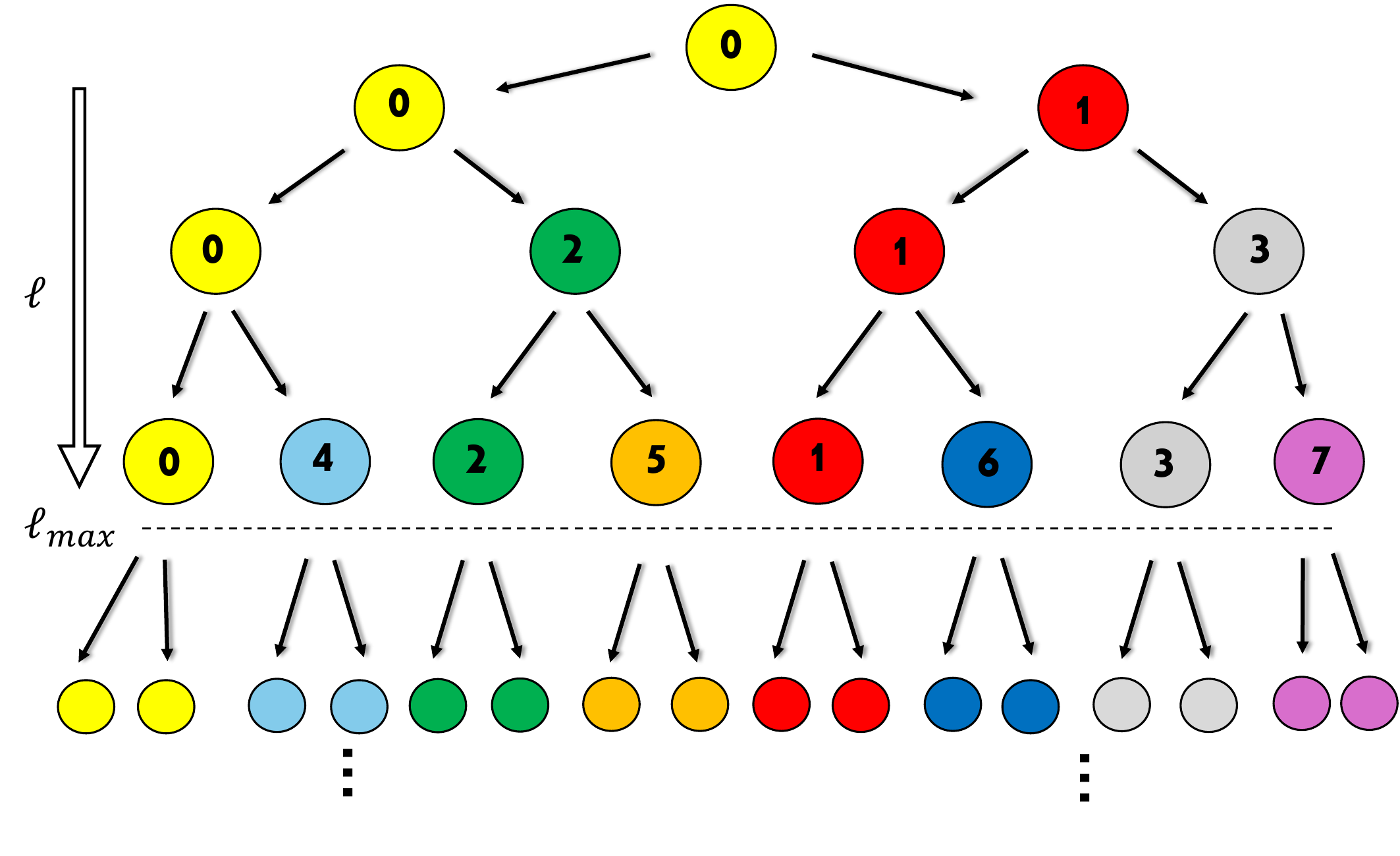}
\caption{$k$-d tree parallelization diagram for a hypothetical case in which $N_\text{CPU} = 8$ and, therefore, $\ell_\text{max} = 3$. Numbers and colors display thread IDs. Each circle represents a set splitting and, hence, the creation of a new tree node (see Fig. \ref{fig:kdtree_diagram}). At $\ell>\ell_\text{max}$, no more threads are spawned, and the splitting workload is handled fully by the corresponding thread, i.e., both the left and right subtrees are built by the same thread.}
\label{fig:parallel_diagram}
\end{figure}

First of all, the program checks whether the tree level is below a given maximum $\ell_\text{max}$, which prevents the algorithm from spawning an arbitrary amount of \texttt{OMP} threads without control. In principle, if all threads are created simultaneously, the maximum level for which we can allow parallelization without exceeding the number of available threads for the \texttt{OMP} job is given by:
\begin{equation}
    \ell_\text{max} = \log_2 (N_\text{CPU})\, ,
\end{equation}
with $N_\text{CPU}$ being the number of available threads allowed to be dedicated to the algorithm. In practice, simultaneity is not accomplished, and a subtree can create a team of threads before the other subtree at the same level, due to having different loads (for instance, having more points to process). Hence, the \texttt{if} condition imposing the $\ell_\text{max}$ condition,  while preventing the code from uncontrolled thread spawning, does not guarantee an ideal parallelization (teams of threads being spawned simultaneously at every level until $\ell_\text{max}$ is reached). In order to illustrate the process, we provide Fig. \ref{fig:parallel_diagram}, depicting the parallelization process for a $N_\text{CPU} = 8$ case. Once the $\ell_\text{max}$ condition for parallelization is met (in this case, $\ell_\text{max} = 3$), all threads perform the subsequent divisions serially, without spawning new companions, as this would lead to having more threads than available.

If the code is allowed to perform parallelization at a given level, the \texttt{parallel} construct is used to define the parallel region, with its corresponding team of threads (two for this application), and the \texttt{single} allows only one thread to execute the subtree building parts inside the \texttt{task} blocks. The \texttt{task} construct, designed to parallelize irregular algorithms such as pointer chasing or recursive algorithms, can be defined within a \texttt{parallel} region, allowing a program to be broken down into smaller, independent units of work that can be executed concurrently by different threads. When a thread encounters the \texttt{task} directive within the \texttt{parallel} region, it executes the code block inside it or defers it to another thread in the team. In this case, the thread entering the \texttt{single} region runs the first \texttt{task} block and defers the second to an available thread from the team\footnote{Note that we do not need to put a \texttt{taskwait} directive, as there is an implicit barrier with the end of the \texttt{single} and \texttt{parallel} regions.}. As it can be seen, at further level, more \texttt{parallel} regions will be defined that are embedded in other \texttt{parallel} regions at lower levels. This algorithm thus produces a structure of nested parallel blocks (\textit{nested parallelism}) which must be allowed with the corresponding environment variable: \texttt{OMP\_SET\_MAX\_ACTIVE\_LEVELS} = $N_\text{nested}$, with $N_\text{nested}$ the maximum level allowed for nested parallelization.

Note that, even with a perfectly balanced tree construction (equal load for all threads), a perfect parallel scaling cannot be accomplished by this approach, as the first splits (which are furthermore the most costly) are not performed with all available threads. It is only after level $\ell_\text{max}$ that full parallelization is obtained. Hence, although our methodology is simple and provides a straightforward parallelization of the $k$-d tree construction, it has its limitations. In Sec. \ref{s:benchmarks}, we examine the tree construction scaling with the number of threads for different CPU architectures.

\subsection{Queries}
\label{s:queries}

\begin{algorithm}[!h]
\caption{$k$-nearest neighbors query: uses a priority queue for finding better neighbors, saved in $\mathcal{K}$.}
\label{alg:knn_query}
\begin{algorithmic}[1]
\Procedure{knn\_query}{$\mathbf{n_0},P,k,\mathcal{K}$}  
\State \textbf{allocate} queue ($\mathcal{Q}$)
\State $N_\mathcal{Q} = 1$
\State $\mathcal{Q}(1) \leftarrow \mathbf{n_0}$
\State
\State \textbf{while} $N_\mathcal{Q} > 0$
\State \quad $\mathbf{n} \leftarrow \mathcal{Q}(1)$
\State \quad $\mathcal{Q}(1) \leftarrow \mathcal{Q}(N_\mathcal{Q})$
\State \quad $N_\mathcal{Q} = N_\mathcal{Q} -1$
\State \quad \textbf{if} $N_\mathcal{Q}>1$
\State \quad \quad \textit{Min Heap}($\mathcal{Q}$)
\State \quad \textbf{end if}
\State
\State \quad \textbf{if} $\mathbf{n}.\text{dist}_{\min} > \mathcal{K}(1).\text{dist}$
\State \quad \quad cycle
\State \quad \textbf{end if}
\State 
\State \quad \textbf{if} $\mathbf{n}$ is a leaf
\State \quad \quad Check points inside
\State \quad \quad Update $\mathcal{K}$ if closer points are found
\State \quad \quad Call \textit{Max Heap}($\mathcal{K}$) for each new point
\State \quad \textbf{else}
\State \quad \quad \textbf{if} $\mathbf{n_L}.\text{dist}_{\min} \leq \mathcal{K}(1).\text{dist}$
\State \quad \quad \quad $N_\mathcal{Q} = N_\mathcal{Q} + 1$
\State \quad \quad \quad $\mathcal{Q}(N_\mathcal{Q}) \leftarrow \mathbf{n_L}$
\State \quad \quad \quad  \textit{Min Heap}($\mathcal{Q}$)
\State \quad \quad \textbf{end if}
\State 
\State \quad \quad \textbf{if} $\mathbf{n_R}.\text{dist}_{\min} \leq \mathcal{K}(1).\text{dist}$
\State \quad \quad \quad $N_\mathcal{Q} = N_\mathcal{Q} + 1$
\State \quad \quad \quad $\mathcal{Q}(N_\mathcal{Q}) \leftarrow \mathbf{n_R}$
\State \quad \quad \quad  \textit{Min Heap}($\mathcal{Q}$)
\State \quad \quad \textbf{end if}
\State \quad \textbf{end if}
\State  \textbf{end while}
\EndProcedure
\end{algorithmic}
\end{algorithm}

\begin{algorithm}[!h]
\caption{Ball query: recursively updates $\mathcal{B}$ with points inside the ball.}
\label{alg:ball_query}
\begin{algorithmic}[1]
\Procedure{ball\_query}{$\mathbf{n},P,R,\mathcal{B}$}
\State Calculate $\text{dist}_{\min}$ and $\text{dist}_{\max}$
\State
\State \textbf{if} $\mathbf{n}.\text{dist}_{\min} > R$
\State \quad  \textbf{return} 
\State \textbf{end if}
\State 
\State \textbf{if} $\mathbf{n}.\text{dist}_{\max} \leq R$
\State \quad add all points within node to $\mathcal{B}$
\State \textbf{end if}
\State 
\State \textbf{if} $\mathbf{n}$ is a leaf
\State \quad Check points inside
\State \quad Add to $\mathcal{B}$ those with distances below $R$
\State \textbf{else}
\State \quad \Call{ball\_query}{$\mathbf{n_L},P,R,\mathcal{B}$}
\State \quad \Call{ball\_query}{$\mathbf{n_R},P,R,\mathcal{B}$}
\State \textbf{end if}
\EndProcedure
\end{algorithmic}
\end{algorithm}

The goal of spending computational resources on building a $k$-d tree structure is to accelerate some calculations that, if done with a brute-force approach, could suppose a considerable bottleneck. In particular, $k$-d trees are an excellent option to accelerate nearest neighbor and range searches using auxiliary subroutines that traverse the tree: the queries. Among all the possibilities, the most useful and widely used implementations are the $k$-nearest neighbors ($k$NN) and ball queries. The first finds the closest $k$ neighbors to a given point, while the second recovers all points within a given distance from the query point. The query benchmarks presented in this manuscript will be restricted to these two modalities, even though we also implement more straightforward queries such as using rectangular (axis-aligned) boxes as ranges.

\paragraph{$k$NN query} $k$-nearest neighbor queries traverse the tree (recursively visit its nodes) to find the $k$-closest neighbors to a given point. The way the nearest neighbors are stored and checked for better candidates, as the tree is traversed, has a major influence on the wall time and scalability of the $k$NN query. After thorough testing, we concluded that the \textit{Max Heap}~\citep{atkinson1986min} structure is the most efficient choice to keep track of the furthest (\textit{max}imum distance) element on the $k$NN list of candidates in order to check for better (closer) neighbors, as the tree is traversed. Shortly, a heap is also a tree-based (binary) data structure that, unlike $k$-d trees (which are designed for spatial indexing), is devoted to store data satisfying a \textit{heap property}. In case this property is the maximum, we call the resulting structure \textit{max heap}, and nodes in the tree are arranged such that the parent node is greater than its children, being the root node, thus, the maximum element. 

Let $\mathcal{K}$ be the set of $k$-nearest neighbors of a query point $P$. In our particular application, we arrange all points in $\mathcal{K}$ in a max heap structure according to the distance to the point $P$, so that the root node is the furthest neighbor. Such a data arrangement is quite convenient for a $k$NN search, as the furthest neighbor candidate can be known in $\mathcal{O}(1)$ time, while inserting a new element on the $\mathcal{K}$ list, satisfying the heap property, has only $\mathcal{O}(\log k)$ complexity. On the other hand, the way the $k$-d tree is traversed is also of utmost importance to quickly retrieve the sought neighbors. To do so, we use a \textit{Min Heap} priority queue to organize all nodes that should be visited. In this regard, the elements of the queue are arranged according to the minimum distance of their bounding box to the target point. Hence, the next element to visit will always be the nearest node to the target found so far. Besides, this minimum distance can also be leveraged to quickly rule out nodes further than the worst candidate in the $\mathcal{K}$ list found so far.

\paragraph{Ball query} In general, range queries traverse the tree to find all points intersecting a given range. In the ball case, a $(D-1)$-sphere centered on a given point is used as range. Contrary to the $k$NN query, there is no need to keep track of the furthest element on the list of results, as all points intersecting the range will be saved. Due to this fact, their querying logic is simpler and, in general, their tree traversals are faster. We use bounding boxes information to quickly check if a node is completely inside or outside the range. In the first case, all points saved at this node child leaves is stored in the resulting list. In the second case, the node and its children are ruled out. In the node is neither contained nor outside the range, its children are visited and the logic is repeated.

For the sake of clarity, Algorithms \ref{alg:knn_query} and \ref{alg:ball_query} provide pseudocodes for our main routines querying the $k$-d tree. To start a query, the code must be supplied by the tree root node $\mathbf{n}_0$, and the query point ($P$). Then, the query routines will visit all child nodes ($\mathbf{n}$) satisfying the necessary conditions to be visited. Regarding the $k$NN query, we keep track of the $k$-nearest neighbor candidates inside the $\mathcal{K}$ (\textit{Max Heap}) list, being the first element of the list the furthest, for fast check and insertion of better candidates. In a similar fashion, all nodes queued for visiting are saved in a \textit{Min Heap} structure $\mathcal{Q}$, such that the first element is the closest node to the target point. In the ball query case, we simply add to the list ($\mathcal{B}$) all points inside the query radius ($R$). The minimum $\text{dist}_{\min}$ (maximum $\text{dist}_{\max}$) distances used to efficiently traverse the tree in both query routines correspond to the minimum (maximum) distances from the target to any point inside the node's bounding box, in the corresponding $D$-dimensional space. 

Once the $k$NN query has found the closest neighbors or the ball query has recovered all points inside a given distance, the algorithm may or may not sort the results, depending on the user's needs. In case that ordering matters, the subroutine uses the \textit{Quicksort} \citep{hoare1961algorithm} algorithm to efficiently ($\mathcal{O}[N\log N]$) sort the final lists. One could possibly think that a \textit{Heapsort}~\citep{williams1964heapsort} algorithm could provide a better performance, as it could take advantage of the already built \textit{Max Heap} structure. Nevertheless, we have checked for different datasets that, generally, \textit{Quicksort} outperforms \textit{Heapsort} by $10-20\%$, even if the latter takes advantage of the already built \textit{Max Heap} structure.

Finally, we want to clarify the fact that our query routines are not explicitly parallelized by design. Instead of parallelizing tree traversal, we seek to process batches of target points concurrently, taking advantage of the CPU-bound nature of the procedures described above.

\subsection{Parameters}
\label{s:parameters}

In physics, $k$-d trees can be useful for performing nearest neighbors or range searches both in real (3D) and phase space (6D), and thus, dimensionality can depend from one case to another. Furthermore, while in some scenarios periodic boundary conditions are not necessary, they can be mandatory for an application on the output of, for instance, a cosmological simulation. Besides, the number of points can be low (e.g. when constructing a tree for galaxies) or large (when using individual simulation particles as points), and low or high precision can be required for floating-point operations. Hence, one may need to use 32-bit or 64-bit integers or floats depending on the application.

Within all this range of possibilities, we devised our $k$-d tree implementation to tackle all of them at once,  allowing the user to specify at compilation time the dimensionality, application of periodic boundary conditions, integer size and floating point arithmetics precision, depending on the need. In this manner, flexibility is maximized, and the $k$-d tree can adapt to all kinds of scenarios. More importantly, since these are pre-compiled options and are taken into account inside the code with conditional compilation directives, they are considered as \texttt{parameters} within the code structure, not as variables, hence allowing the \texttt{Fortran} compiler to still produce aggressive optimizations, avoiding speed penalties.

\subsection{Python bindings}
\label{s:bindings}

In general, \texttt{Python} has a much wider range of applications than \texttt{Fortran}, as it is an interpreted general-purpose programming language. In that regard, so as to significantly expand \cosmokdtree{}'s potential applications, we provide a \texttt{Python} module compiled with \texttt{F2PY} \cite{harris2020array} offering a similar performance to the \texttt{Fortran} module that can be easily imported and accommodated into new or preexisting \texttt{Python} codes.

\section{Benchmarks}
\label{s:benchmarks}

In this section, we present several tests to benchmark the tree construction and query performances in uniform and inhomogeneous point distributions both for the \texttt{Fortran} and \texttt{Python} modules. A comparison with \scipy{}'s \citep{virtanen2020scipy}, \coretran{}'s \citep{coretran} and \torchkdtree{}'s \citep[GPU,][]{torch_kdtree} $k$-d trees is also presented. The first two libraries are widely used by the \texttt{Python} and \texttt{Fortran} communities, respectively, having been designed for accelerating and facilitating data analysis for scientific applications. On the other hand, \torchkdtree{} is a public implementation \citep{torch_kdtree} of a $k$-d tree in \texttt{CUDA} \citep{cuda} that carries out the construction phase on the GPU and queries on the CPU. The default CPU and memory configuration for running all benchmarks (unless otherwise specified) consists of an AMD Ryzen Threadripper PRO 5965WX (24 cores) with 256 GB of RAM (DDR4). We carry out all GPU benchmarks on a NVIDIA GeForce RTX 5090 with 21760 \texttt{CUDA} cores and 32 GB of VRAM (GDDR7)\footnote{This GPU belongs to a substantially newer hardware generation than the CPU.}. In Table \ref{tab:versions} we provide the software versions used for obtaining our results.

\begin{table}[h!]
\centering
\caption{Software versions}
\label{tab:versions}
\vspace{0.2cm}
\begin{tabular}{|c|c|c|c|c|c|c|}
\hline
\texttt{Python} & \texttt{SciPy} & \texttt{gfortran} & \texttt{OpenMP} & \texttt{Coretran} & \texttt{CUDA} & \texttt{TORCH\_KDTREE}\\
 \hline
 3.11.5 & 1.11.1 & 11.4.0 & 4.5 & 1.0.2 & 13.2 & 63 (commit)\\
 \hline
\end{tabular}
\end{table}

The following \texttt{gfortran} flags where used to compile the code: \texttt{-O3 -fopenmp -ftree-vectorize -funroll-loops -march=native -mcmodel=medium}

\subsection{Mock tests}
\label{s:tests}

\begin{figure}[h!]
\centering 
\includegraphics[width=0.5
\linewidth]{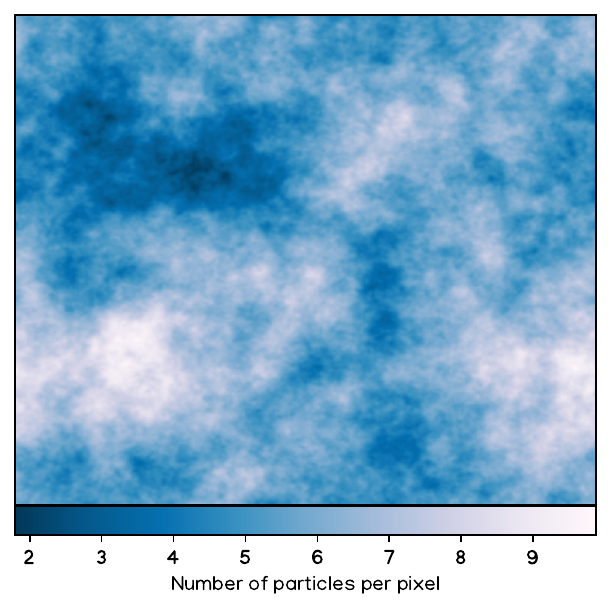}
\caption{2D slice of one of the inhomogeneous distributions of particles used to test the $k$-d tree. In this case $N = 10^7$ and the number of pixels corresponds to $128^3$.}
\label{fig:inhomogeneous_distribution}
\end{figure}

\begin{figure}[h!]
\centering 
\includegraphics[width=0.65\linewidth]{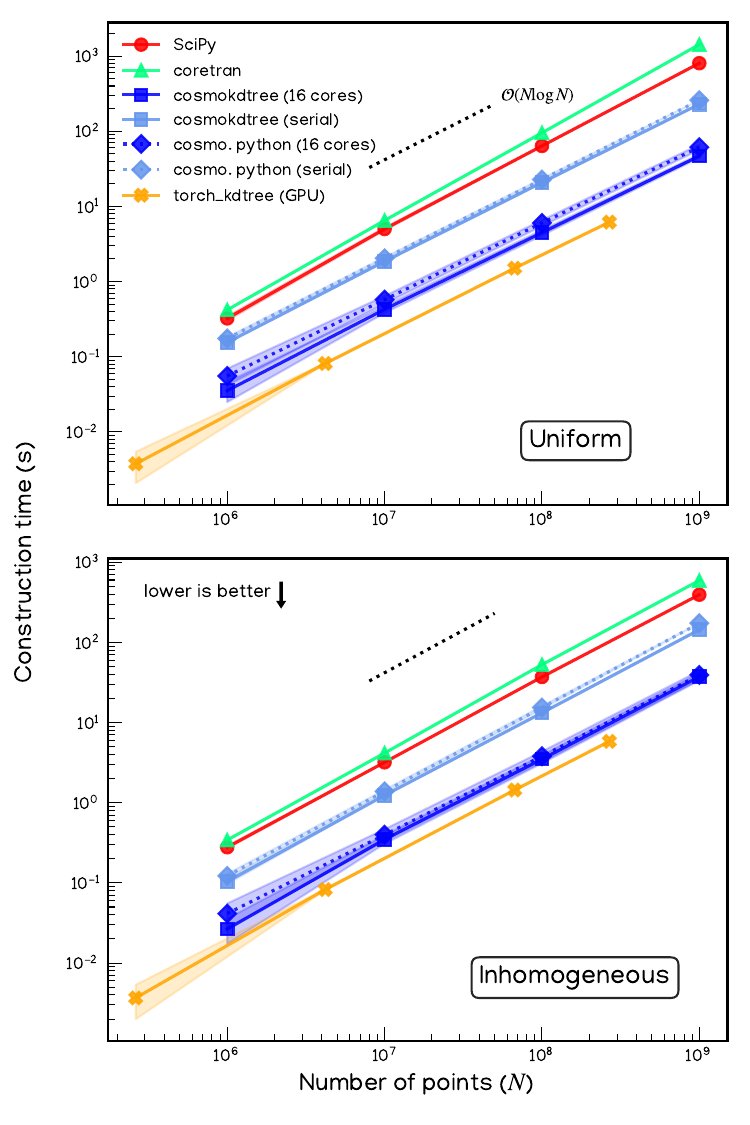}
\caption{Tree construction (3D) benchmark for \scipy{} (circles), \coretran{} (triangles), \torchkdtree{} (crosses) and our implementation (lighter for 1 thread, darker for 16 threads, diamonds and dotted lines for the \texttt{Python} module and squares and continuous lines for the \texttt{Fortran}-only version). Shaded regions display 1 standard deviation. \textit{Upper panel} displays the construction time for a set of points uniformly distributed in space, while the \textit{lower panel} shows the inhomogeneous case. The $\mathcal{O}(N\log N)$ complexity is plotted for reference.}
\label{fig:construction}
\end{figure}

\begin{figure}[h!]
\centering 
\includegraphics[width=0.7\linewidth]{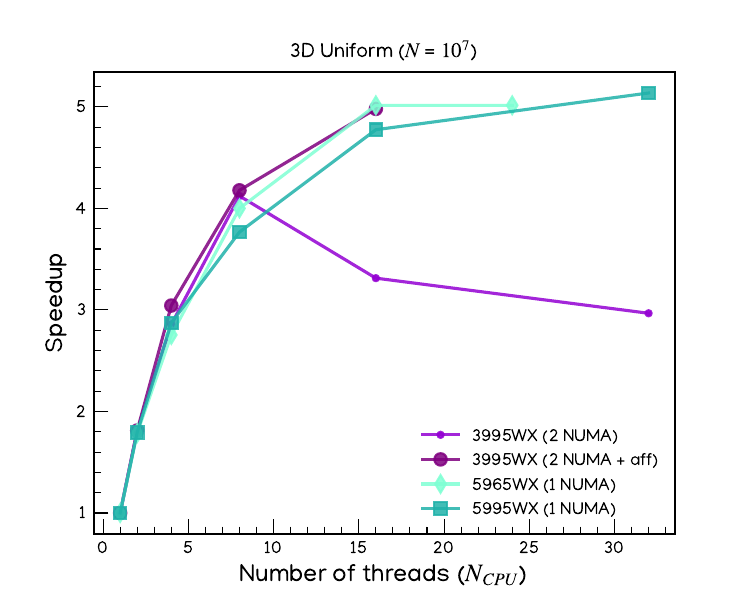}
\caption{Tree construction speedup against the number of available threads for four CPU configurations. All runs were performed over a uniform distribution of $10^7$ points in 3D.}
\label{fig:construction_cpu}
\end{figure}

\begin{figure}[h!]
\centering 
\includegraphics[width=0.7\linewidth]{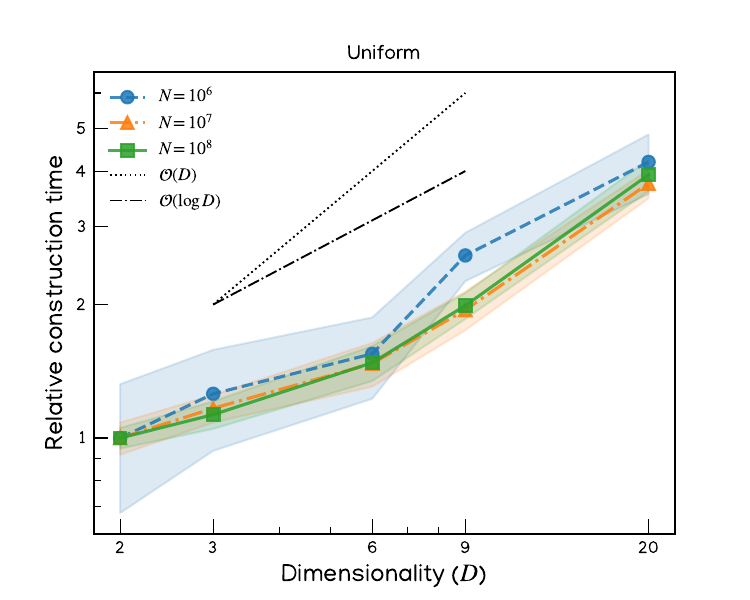}
\caption{Tree construction time (relative to the 2D case) against dimensionality for $N=10^6$ (circles), $N=10^7$ (triangles) and $N=10^7$ (squares). Shaded regions display 1 standard deviation. The distribution of points is uniform in all cases. The $\mathcal{O}(N)$ and $\mathcal{O}(\log N)$ scalings are displayed for reference.}
\label{fig:construction_dim}
\end{figure}

We have produced two complementary setups for the tests: a uniform and an inhomogeneous particle distribution. This distinction is critical, since the efficiency of space-partitioning trees is hugely impacted by the homogeneity of the input data or lack thereof \citep{maneewongvatana1999analysis}. When the data is clustered or sparse, the resulting tree can be unbalanced and excessively refined in some regions, in such a way that increasing depth leads to poorer performance. In both cases, all particles belong to a hypercube (in $D$ dimensions) of sidelength $L$.

\paragraph{Uniform particle distribution} The distribution of particles can be straightforwardly obtained by drawing the values of each coordinate $x_i$ from a uniform, $\mathcal{U}(0,1)$, distribution, mapped to the $(0, L)$ range.

\paragraph{Inhomogeneous particle distribution} In order to recreate a more realistic situation, with $N$ particles non-uniformly distributed, we first generate an inhomogeneous density field on a grid with $N_x \sim \lfloor \sqrt[3]{N} \rfloor$ cells along each direction, so that the mean interparticle separation is similar to the grid spacing. We then construct the density field in Fourier space by sampling from a Gaussian random field with spectral index $n_s=-2$, which has been empirically chosen so as to generate a density field with sufficiently large density perturbations. Finally, the particle distribution is obtained by assigning particles to cells proportionally to their density, adding a uniformly sampled, random shift of the particle positions inside each cell. In Fig. \ref{fig:inhomogeneous_distribution}, we display a 2D slice of an inhomogeneous distribution of particles produced with this method.

\subsection{Construction time}
\label{s:construction_time}

The first point we assess in order to demonstrate \cosmokdtree{}'s performance is the tree construction time. Below, we study this issue as a function of the number of input points, their spatial distribution, the number of available threads, CPU architecture and dimensionality of the input space.

Figure \ref{fig:construction} presents the tree construction benchmark for \cosmokdtree{} (light blue for single-threaded and dark blue for parallel build with 16 threads), \scipy{} (red), \coretran{} (green) and \torchkdtree{} (yellow). \cosmokdtree{}'s python module performance is also displayed with dotted lines. In all cases, construction routines are run 5 times to calculate the mean wall time standard deviation. Both the uniform and inhomogeneous construction times are presented for points distributed in a 3D space with varying $N$. In both scenarios, we can observe that the three algorithms follow the $\mathcal{O}(N\log N)$ expected behaviour. Both \cosmokdtree{} (CPU) and \torchkdtree{} (GPU) parallel implementations consistently outperform sequential constructions by over an order of magnitude ($\sim 10-20$ factor), with the latter achieving the best (fastest) performance among the tests. Among serial runs, \cosmokdtree{}'s stands as the fastest, with construction times $\sim3$ (4) times smaller than \scipy{}'s (\coretran{}'s) implementations. On the other hand, the \cosmokdtree{}'s \texttt{Python} module follows the performance of the \texttt{Fortran}-only version with a small overhead added due to the wrapper functions needed to transfer data between the Python interpreter and the \cosmokdtree{} binary. Due to the limited VRAM available for our GPU, running \torchkdtree{} on $N \gtrsim 3\times10^{8}$ was not possible. Besides, in this particular case, $N$ had to be restricted to powers of $2$ due to \torchkdtree{}'s code structure limitations that are inherited from GPU architecture. In the tested range of $N$, \torchkdtree{} attains the lowest tree construction times when run on a high-end consumer GPU (RTX 5090). In the broader range of values of $N$ in which \cosmokdtree{} can be applied, it stands as the fastest CPU-based version when running the tests on an AMD Ryzen Threadripper PRO 5965WX with 16 threads, building the tree just a factor of $\sim 2$ slower than \torchkdtree{}.

To put these results into context, we can compare with the results of the GPU-based algorithm of \citep{wald2022gpu} tested on GPUs belonging to previous hardware generations than ours. The best times are achieved by an NVIDIA GeForce RTX 3090Ti with 10752 \texttt{CUDA} cores and 24 GB of VRAM (GDDR6X): for $N = 10^6, 10^7$ and $10^8$ their implementation takes $44.4$ ms, $424$   ms and $5.3$ s to build the tree, thus being on average $40\%$ slower than \cosmokdtree{} being run on our configuration. All in all, a cautious conclusion should be extracted from the construction test presented in Fig. \ref{fig:construction} and the results presented in \citep{wald2022gpu}: \cosmokdtree{} is able to significantly outperform other existing CPU implementations in the construction phase, being able to match or slightly fall behind some GPU implementations under high-end consumer hardware specifications. We do not further delve into the comparison with GPUs, as this would require a different hardware setting focused on dedicated data-centre accelerators (e.g., NVIDIA H100), which falls beyond the scope of this work. Rather than that, our aim here is to characterize performance in a more general-purpose environment, representative of typical analysis workstation hardware, and comparable to our choice of CPU.

From the previous analysis, it can be seen that our parallelization for the $k$-d tree construction yields a substantial speedup, which can considerably reduce the wall time in computationally demanding tasks. Nevertheless, as we already pointed out in Sec. \ref{s:parallelization}, although our approach is simple to implement, it also has its limitations. The inherent sequential structure of the algorithm at the first levels cuts down the speed gains since increasing the number of threads does not improve performance in the first partitioning steps. So as to illustrate this behavior, we provide Fig.~\ref{fig:construction_cpu}, which displays the tree construction speedup against the number of available threads for four different CPU configurations, all being part of the AMD Ryzen Threadripper PRO processor series. We run the code with the 3995WX, 5965WX and 5995WX units, with the former being launched in two different configurations: the first does not specify a thread affinity for the \texttt{OMP} task, while the second does. As it can be seen, the speedup obtained is the same for all cases, except for the 3995WX without affinity, for which a poor behavior is obtained when surpassing 8 threads. In general, the speedup flattens at a value of $5$ for $N_\text{CPU} = 16$, from which little speed gains are obtained.

The undesired speedup performance for the 3995WX run (without thread affinity) is caused by its architecture, which is different from the more modern 5965WX and 5995WX. While the first has two different Non-Uniform Memory Access (\texttt{NUMA})\footnote{The \texttt{NUMA} architecture divides memory and CPUs into \textit{nodes}. Each node has its local memory and CPU cores. Accessing local memory is fast, but accessing remote memory (on another node) incurs latency and potential bandwidth penalties.} nodes, the other two have only one. The memory access between threads inside the same \texttt{NUMA} node is fast, while some overhead is added if communication between two different nodes is required. This is the reason why the 3995WX run with thread affinity is faster than the one without it: it forces all threads to belong to the same \texttt{NUMA} node, ensuring optimal memory transference between them. It becomes clear that the runs should be \texttt{NUMA}-aware in order to ensure the best performance.

Another important point to assess is the tree construction time scaling with the number of dimensions. Figure \ref{fig:construction_dim} displays this scaling for different numbers of input points. The construction routine was run 10 times for each $N$ and $D$ to get the mean wall time and standard deviation. We have tested that the high variability ($\sim30\%$ in the worst case) that can be observed for the $N = 10^6$ case is due to how OpenMP is handling memory and thread-spawning in the nested parallelism (see Sec. \ref{s:parallelization}). As can be observed, although constructing the tree gets more costly for increasing $D$, the scaling is limited to $\mathcal{O}(\log D)$ within $D \in (2, 20)$ and $N \in (10^6, 10^8)$. This result expands \cosmokdtree{}'s applicability beyond the usual $D=2,3$ dimensionalities, hence increasing its potential use cases.

\begin{figure*}[h!]
\centering 
\includegraphics[width=1\linewidth]{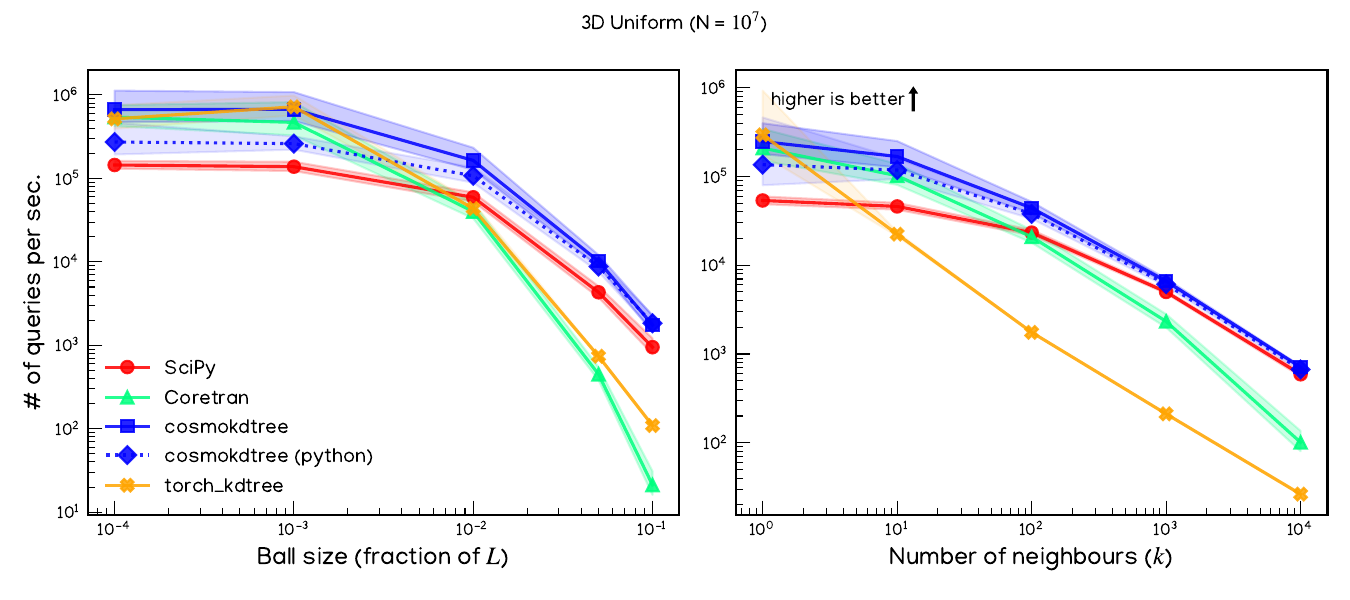}
\caption{Number of $k$NN (\textit{right}) and ball (\textit{left}) queries per second (throughput) as a function of search size: number of neighbors for the former and ball radius for the latter. The \scipy{}'s (circles), \coretran{}'s (triangles) and \torchkdtree{} (crosses) throughput scalings are presented together with \cosmokdtree{}'s results (squares and diamonds for \texttt{Fortran} and \texttt{Python} modules, respectively). Shaded regions display 1 standard deviation. An identical input was provided to the three algorithms, consisting of $N=10^7$ points uniformly distributed in a 3D space. For this test, queries are run serially.}
\label{fig:query_uniform}
\end{figure*}

\begin{figure*}[h!]
\centering 
\includegraphics[width=1\linewidth]{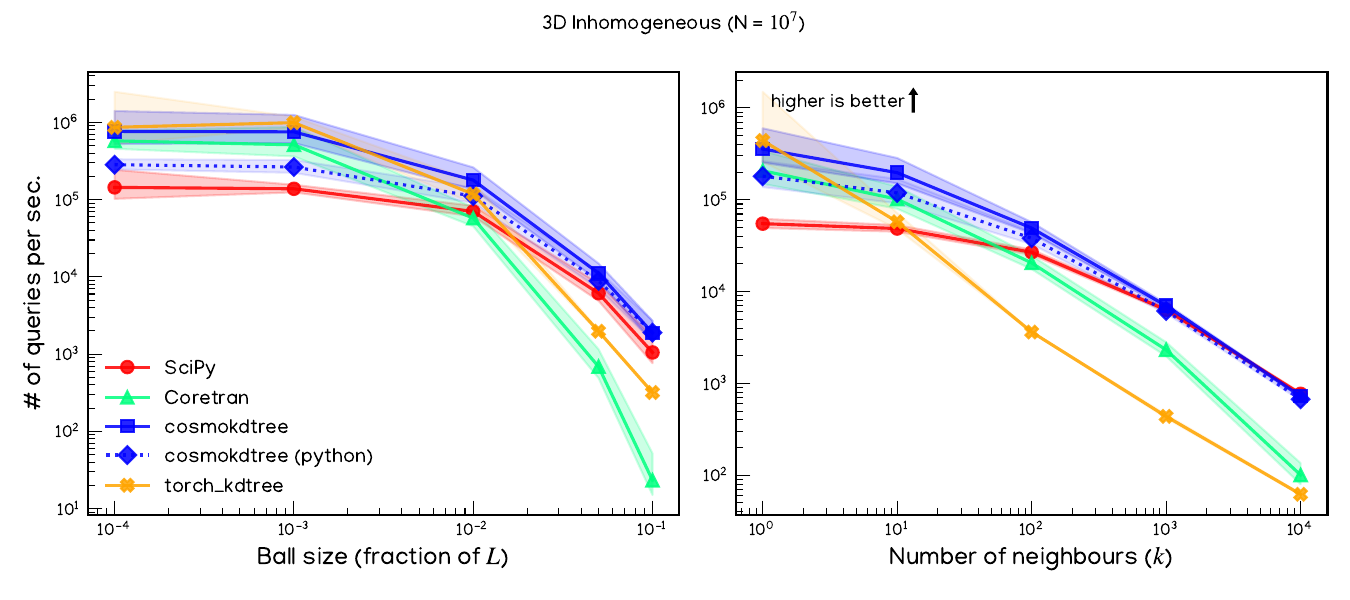}
\caption{Same as Fig.~\ref{fig:query_uniform} but for an inhomogeneous distribution of points.}
\label{fig:query_anisotropic}
\end{figure*}

\begin{figure}[h!]
\centering 
\includegraphics[width=0.7\linewidth]{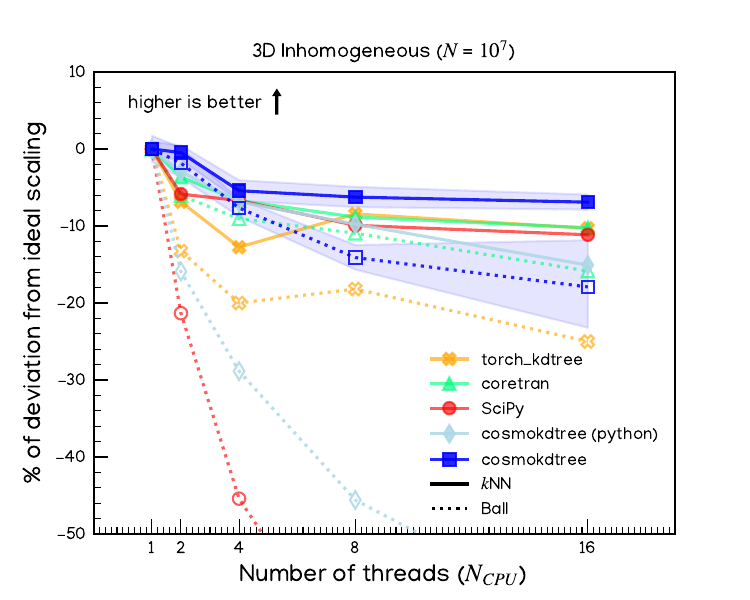}
\caption{Percentage of deviation from ideal scaling as a function of the number of threads both for the $k$NN (solid lines) and ball queries (dotted lines) with $k = 100$ and $R = 0.01 L$, respectively. In all runs, the number of queries calls distributed across the available threads was set to $N_\text{query} = 10^5$. All runs were performed over an inhomogeneous distribution of $10^7$ points in 3D. Data is displayed for \scipy{} (circles), \coretran{} (triangles), \torchkdtree{} (crosses),  \cosmokdtree{} (squares) and \cosmokdtree{}'s \texttt{Python} module (diamonds). For the sake of clarity, only \cosmokdtree{}'s standard deviation (shaded region) is displayed.}
\label{fig:query_parallel}
\end{figure}

\begin{figure*}[h!]
\centering 
\includegraphics[width=1\linewidth]{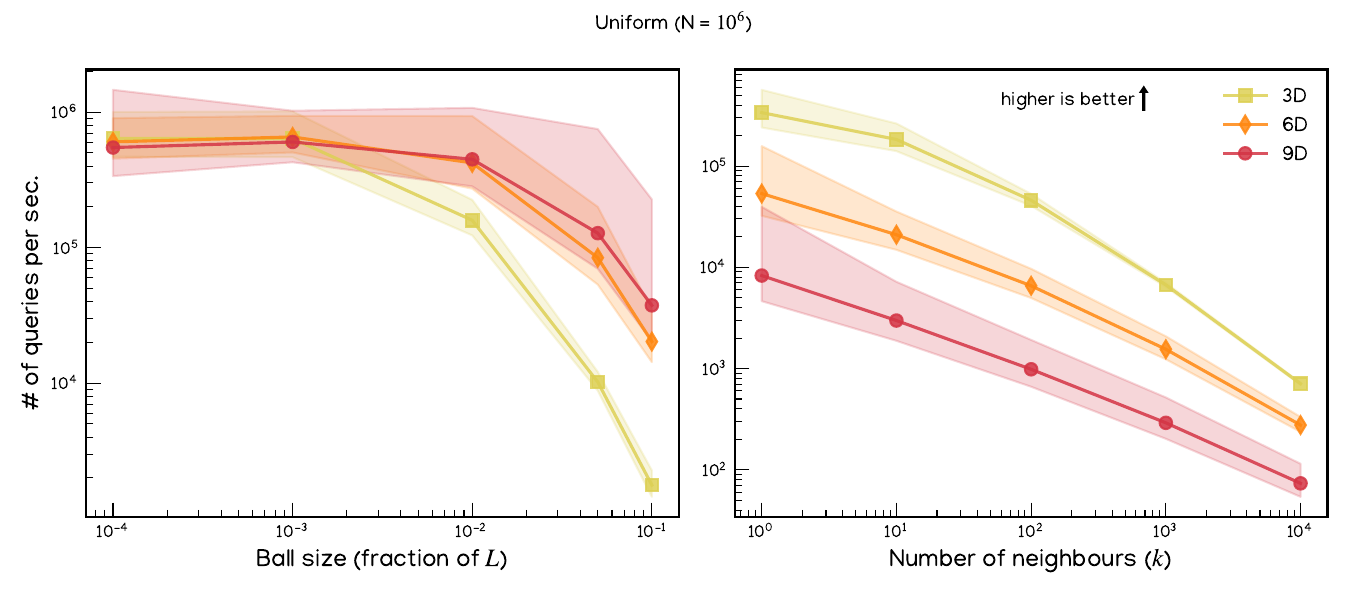}
\caption{Same as Fig.~\ref{fig:query_uniform} but instead of comparing different $k$-d tree implementations, we compare \cosmokdtree{}'s query performance on a dimensionality basis: 3D (squares), 6D (diamonds) and 9D (circles).}
\label{fig:query_dimensionality}
\end{figure*}

\subsection{Query time}
\label{s:query_time}

After testing the tree construction efficiency, the most important point to evaluate corresponds to the queries. An efficient tree construction routine may be thrown away by poor performance in the querying phase. To delve into this, we study the performance of the $k$NN and ball queries as a function of search size\footnote{Search size refers to the number of nearest neighbors ($k$) in the context of the $k$NN query and the ball radius ($R$) in the context of the ball query.} and dimensionality.

Figure \ref{fig:query_uniform} and \ref{fig:query_anisotropic} display the number of queries per second (throughput) against search size, both for the $k$NN and ball modalities, for a uniform and inhomogeneous distribution of points in a 3D space, respectively. For comparison, the \scipy{}, \coretran{} and \torchkdtree{} scalings are also plotted. For \torchkdtree{}'s case note that, although the construction phase is GPU-based, queries are carried out on the CPU. In all cases, for each search size, the query routines were run $10\,000$ times to obtain the mean wall time and standard deviation. Dispersion is higher for smaller times as they are more affected by CPU execution variability. Also, it is worth emphasizing, the fact that in these figures we are benchmarking serial query throughput, since parallel execution of queries is tested below (Fig. \ref{fig:query_parallel}). Regarding $k$NN queries, \cosmokdtree{} achieves the best performance across all implementations, consistently outperforming the other for $10 \leq k \leq 1000$. For higher ($k \sim 10^4$) [lower, $k \sim 1-10$] values of $k$ it is statistically as efficient as \scipy{} [\coretran{}]. Similar conclusions can be drawn for the case of the ball query, where \cosmokdtree{} attains the lowest query times across all implementations except at the lowest ball radii, where instead it is statistically compatible with \torchkdtree{}. 

A final comment must be delivered regarding \texttt{Python} implementations. Both \scipy{} and \cosmokdtree{}'s \texttt{Python} modules suffer from overhead due to the interface between \texttt{C++} and \texttt{Fortran} routines (respectively) and \texttt{Python}. This interface the overall query time for small sizes. It can be seen, however, that \cosmokdtree{}'s module incurs less overhead than \scipy{}'s, increasing the overall query efficiencies at all times but, especially, at lower search sizes.

Contrary to the tree construction phase, query routines do not traverse the tree concurrently and, thus, a single query call is always executed serially (see Algorithms \ref{alg:knn_query} and \ref{alg:ball_query}). Far from being a drawback, this opens the door to executing multiple query calls in parallel, as each thread performs its own tree traversal (spatial search) asynchronously. The resulting speedup depends on the query routines memory-management efficiency. To test \cosmokdtree{}'s query parallel scalability, we run $N_\text{query} = 10^5$ query calls in parallel for the $k$NN and ball queries with $k = 100$ and $R = 0.01 L$, respectively, varying the number of threads dedicated to perform the query calls concurrently. The input dataset consists of $N = 10^7$ points inhomogeneously distributed in a cube. We carry out similar tests with \scipy{}, \coretran{} and \torchkdtree{}\footnote{In \torchkdtree{}'s case, $N = 2^{23} \approx 8\times10^6$ is used instead of $10^7$ so as to fit the code and GPU architecture.} and present the results in Fig. \ref{fig:query_parallel}, where the percentual deviation from ideal scaling is plotted against the number of threads used for parallelization. For a given parallel speedup $s \equiv t_\mathrm{serial} / t_\mathrm{parallel}$ using $N_\text{CPU}$ threads, we define the deviation from ideal scaling as $s/N_\text{CPU} - 1$.

In general, $k$NN queries obtain the best parallel speedup, as they only deviate by $\sim 15\%$ at most from ideal scaling. The best performance is achieved by \cosmokdtree{}'s $k$NN routine. On the other hand, ball queries are less efficiently parallelised, with deviations ranging from $15\%$ to $60\%$. In particular, while \cosmokdtree{}'s, \coretran{}'s and \torchkdtree{}'s ball query routines are just slightly less efficient than their corresponding $k$NN search, \scipy{}'s and \cosmokdtree{}'s \texttt{Python} module (but not \cosmokdtree{}'s pure \texttt{Fortran} implementation) yield significantly slower results. The difference in parallel performance between the different types of queries can be attributed to the fact that in $k$NN searches, the length of the resulting array is known before traversing the tree ($k$), while in the ball case this is not possible. This distinction is key, as $k$NN queries are thus more CPU-bound and, ultimately, efficiently parallelized.

Another aspect that needs careful attention regards how query performance degrades as we increase dimensionality. Indeed, as we have already presented in Sec. \ref{fig:construction_dim}, the tree construction time increases with increasing number of dimensions. However, although this feature is certainly undesired, its impact can be relegated to a second plane when the query performance scaling with the number of dimensions is taken into account. In Fig. \ref{fig:query_dimensionality}, we present the number of queries per second attained by \cosmokdtree{} as a function of search size, for three different dimensionalities. We run the query routine 10.000 times for each search size and dimensionality to accurately quantify the mean wall time and standard deviation. The performance penalty for the $k$NN query is substantial at all $k$: results for the 6D case are an order of magnitude slower than the 3D ones, whilst the 9D results are also an order of magnitude slower than the 6D case. On the contrary, ball queries increase their performance for higher $D$. The observed behavior for $k$NN searches occurs because of the \textit{curse of dimensionality}: for a fixed number of points, if dimensionality is sufficiently increased, the volume between points becomes so large that the dataset becomes sparse and very dissimilar, preventing $k$-d trees from producing optimal space partitions. For a comprehensive review on the effects of the \textit{curse of dimensionality} on nearest-neighbor searches, see \citep{marimont1979nearest}. On the other hand, the performance increase for ball searches is explained by the fact that, as points become more and more sparse when $D$ is increased, bounding box pruning (see Algorithm \ref{alg:ball_query}) becomes more efficient. In Cosmology, as well as in most physical applications, needing more than 6-dimensional spaces (e.g., phase space) is unusual and, thus, in general, these issues can be safely neglected.

\subsection{Memory usage}

\begin{figure}[h!]
\centering 
\includegraphics[width=0.7\linewidth]{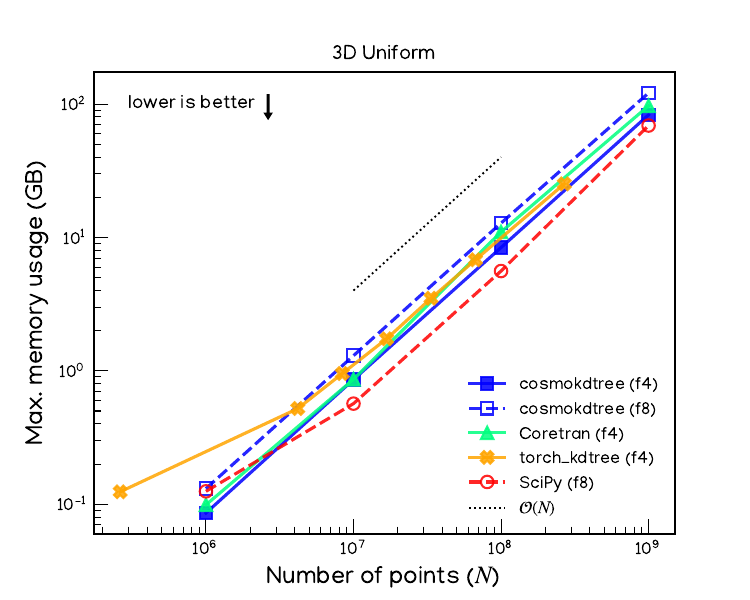}
\caption{Maximum memory usage when constructing the $k$-d tree for \cosmokdtree{} (squares), \scipy{} (circles), \coretran{} (triangles) and \torchkdtree{} (crosses). A dashed line is used to display \cosmokdtree{}'s and \scipy's memory usage with double-precision floats (f8). The $\mathcal{O}(N)$ scaling is displayed for reference.}
\label{fig:memory_footprint}
\end{figure}

Although a $k$-d tree package should minimize tree construction and query times, it is also important to keep memory usage low for scalability. In this direction, we measured the peak resident memory of \cosmokdtree{}'s (both in single and double-precision floats), \scipy{}'s, \coretran{}'s and \torchkdtree{}'s tree construction routines against the number of input points. Results are displayed in Fig.  \ref{fig:memory_footprint}. All implementations display a $\mathcal{O}(N)$ memory usage scaling. Single-precision float versions display a similar memory usage with \torchkdtree{} having the highest for small $N$ and \coretran{} the highest for larger $N$. Among these, \cosmokdtree{} shows the overall lowest memory consumption. Regarding double-precision float trees, \scipy{} displays the best performance, with \cosmokdtree{} using roughly two times more memory.

\cosmokdtree{} was devised to maximize query and tree construction efficiencies. To achieve the former, the tree structure has to save extra information (namely, node bounding boxes), in what could be considered a tradeoff between performance and memory usage. Furthermore, the $k$-d tree is not build in-place: we copy the original input data and perform partitioning and reordering on that auxiliary array that, at the end of tree construction, is deallocated. In addition, leaf nodes save a copy of the portion of input data they contain. While extra memory usage is also added following these steps (hence explaining the difference with \scipy{}) we do it in purpose to maximize efficiency and preserve thread-safety and memory locality. The resulting $k$-d tree is self-contained and does not modify/depend on external arrays, thus becoming and ideal tool to potentially embed in wider codes such as cosmological simulations or halo finders needing spatial indexing acceleration.

\section{Applications}
\label{s:application}

After having displayed \cosmokdtree{}'s performance with a set of different tests highlighting its efficiency when dealing with varying sizes of input datasets, dimensions or parallelization options, in this section we present two applications for which this $k$-d tree implementation could result remarkably helpful saving computational resources. 

The first case corresponds to the friends-of-friends clustering and grouping algorithm, widely used in astrophysics for structure identification purposes (see, e.g., \citet{knebe2011haloes}). The second application corresponds to the problem of extracting a continuous, grid description from a particle representation of the physical quantities (or variables in general). The $k$NN query can be leveraged to accelerate this costly process. In this direction, data analysis and visualization packages \citep{price2007splash} or simulation post-processing tools \citep{valles2024vortex, monllor2025avism} could also take advantage of \cosmokdtree{}.


\subsection{Friends-of-friends, clustering and grouping algorithms}
\label{s:fof}

The \textit{friends-of-friends} (FoF) grouping algorithm is a widely used tool in several scientific fields (e.g., cosmology) where identifying groups or clusters in data is important. It works as follows: given a set of points in an arbitrary metric space $X$, each point is linked to its neighbors within a distance $l_\text{FoF}$ (its direct \textit{friends}) and, by recursion, to all points connected to its \textit{friends} (hence, the term  \textit{friends-of-friends} ).

This technique is extensively applied to the \textit{halo-finding} problem (see \citet{knebe2011haloes} for an explicit definition of this issue and comparison of different methodologies), which refers to the problem of identifying gravitationally bound structures (dark matter haloes) and their substructures, within the output of cosmological simulations. Among all halo finders using the FoF approach, some of the most refined and commonly used in numerical cosmology are \textsc{subfind} \citep{springel2001populating, dolag2009substructures}, in 3D configuration space, or \textsc{rockstar} \citep{behroozi2012rockstar}, in 6D phase space.

Finding all neighbors within a given distance $l_\text{FoF}$ for all points in a given dataset can be costly, as the number of calculations using a \textit{brute-force} approach scales as $\mathcal{O}(N^2)$ and modern dataset sizes are vast (e.g., $N \sim 10^{9-11}$ for modern cosmological simulations, see \citet{maksimova2021abacussummit}). Several techniques can be applied to accelerate this computation: creating an auxiliary grid for linking particles in close cells, sorting particles by the $x$ coordinate \citep{valles2022halo}, using space-partitioning tree structures, etc. Among all options, the $k$-d tree represents one of the best, as it is cheap to build and it is able to reduce the computational cost to an $\mathcal{O}(N \log N)$ complexity, accelerating the FoF groups identification considerably.

For the purpose of exemplifying our $k$-d tree capabilities to accelerate the FoF group search, we have implemented the FoF algorithm in \texttt{Fortran} and designed three different neighbor search kernels to look for the friends of a given point. One uses the brute-force approach, another takes advantage of a uniform auxiliary grid to restrict the neighbor search to particles in close cells, and the last one uses \cosmokdtree{}. These three different versions of the FoF algorithm are benchmarked using a test consisting of $N_\text{group} = 300$ groups uniformly distributed on a $L = 1$ cubic box (3D real space). The spatial distribution of points inside each group is sampled from a 3D isotropic Gaussian distribution with $\sigma = 0.1$. A background set of points is also placed, using a uniform distribution across the whole domain. The number of points in groups and background is set to $99\%$ and $1\%$ of the total ($N$), respectively. The resulting distribution would resemble, to some extent, a distribution of clusters of galaxies inside a cosmological volume, which is a prototypical example for clustered data. Since $N$ can vary freely but the underlying spatial distribution is the same (only varying the number of elements sampling it), we can study the FoF algorithm performance purely as a function of the number of input points, minimizing the impact of how these points are spatially distributed. Lastly, to obtain an optimal behavior of the FoF algorithm, a proper linking length must be chosen. The most common method sets $l_\text{FoF}$ to a fraction $b$ of the mean interparticle  separation $\langle l\rangle = (V/N)^{1/3} = 1/\langle \varrho \rangle ^{1/3}$, so that $l_\text{FoF} = b\, \langle l \rangle$, being $V$ the total volume and $\langle \varrho \rangle$ the mean particle number density. The parameter $b$ determines the approximate value of the isodensity contour ($\varrho_\text{FoF}$) delimiting the structure: $\frac{4}{3} \pi l_\text{FoF}^3 \varrho_\text{FoF} \sim 1$. Joining these expressions, we can get an equation for $b$:
\begin{equation}
b = \bigg( \frac{3}{4\pi} \frac{\langle \varrho\rangle}{\varrho_\text{FoF}} \bigg)^{1/3}\, .    
\end{equation}
Using this formula on the set of groups described before, we can easily obtain $b = 0.03$ as the proper value for finding the mock groups created to test the algorithm.

\begin{figure}[h!]
\centering 
\includegraphics[width=0.7
\linewidth]{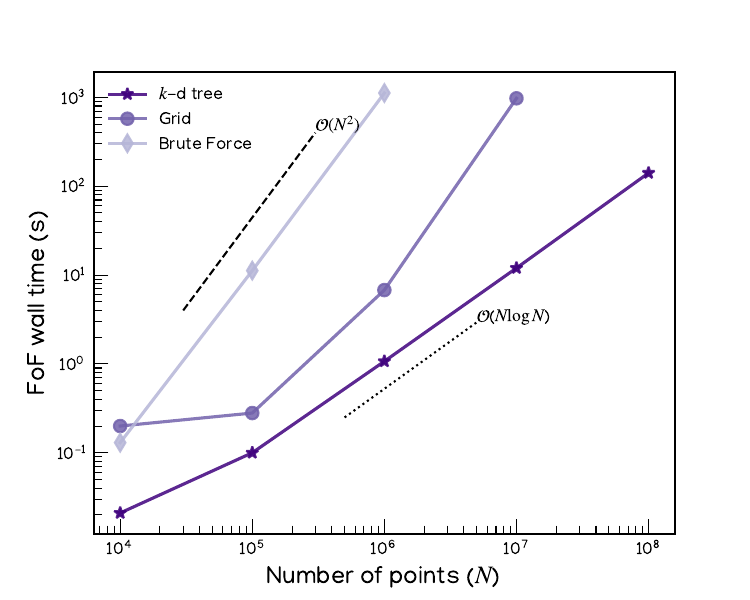}
\caption{Time to identify all friends-of-friends (FoF) groups as a function of the number of input points. Results for the brute-force (diamonds), auxiliary grid (circles) and $k$-d tree (stars) approaches to find all neighbors within a linking length distance are displayed. The linking length for each dataset is set to $l_\text{FoF} = 0.03 \,\langle l \rangle$, being $\langle l \rangle$ the mean interparticle separation. As a reference, the $\mathcal{O}(N^2)$ and $\mathcal{O}(N \log N)$ time complexities are also displayed.}
\label{fig:fof}
\end{figure}

Figure \ref{fig:fof} shows the FoF wall time to find all groups as a function of the dataset size. All FoF runs were performed without parallelizing the group identification. The time scaling is displayed for the three different kernels used to find all neighbors within a linking length distance: brute force (diamonds), auxiliary grid (circles) and our $k$-d tree (stars). Both the auxiliary grid and $k$-d tree implementations outperform the brute-force approach, significantly reducing the overall wall time to get the underlying FoF groups. Nevertheless, while our particular implementation of the grid technique is not able to avoid the undesired $\mathcal{O}(N^2)$ scaling, the one accelerated by the $k$-d tree is able to maintain a more optimal $\mathcal{O}(N \log N)$ behavior, vastly outperforming the other two at large $N$ and, hence, dramatically reducing computational costs in this regime. Such results suggest that using the $k$-d tree technique to look for neighbor points in the FoF algorithm is one of the best options to tackle large input datasets, such as those produced by modern simulations.

\subsection{Smooth particle-to-grid assignment}
\label{s:part_smooth}

Although the use of a set of particles to describe the evolution and spatial distribution of a set of physical quantities can be very convenient in some cases (a good example would be the Smoothed-Particle Hydrodynamics approach \citep[SPH,][]{monaghan2005smoothed} used in many cosmological codes \citep{springel2001gadget}), sometimes it is necessary to transform the particle representation of the physical quantities to a grid (Eulerian) description. For instance, in some cases, it could be desirable to carry out some kind of calculations for which using point particles as numerical tracers is not the best option. On the other hand, regarding the illustration of physical quantities, the continuous description always provides smoother and clearer representations than the clumpy particle plots and, thus, it is frequently sought to depict numerical results, especially for volume renderings (e.g. see \citet{price2007splash}).

Several interpolation schemes exist to spread particle quantities onto a grid \citep{hockney2021computer}. The simplest technique, known as \textit{nearest grid point} (NGP), assigns the particle quantities to the nearest cell. Clearly, this scheme will sample the physical quantities too crudely and is rarely used. Better assignment techniques spread the particle variables over the $2^{D}, 3^{D}, \dots$ closest cells, with $D$ the dimensionality of the grid. The $2^D$ case is known as the \textit{cloud in cell} (CIC) assignment, while if the variables are spread over the $3^D$ closest cells, the \textit{triangular shape cloud} scheme (TSC) is used. The former yields continuous fields, while the latter produces both continuous and differentiable results. These series of schemes are easy to implement and computationally cheap ($\mathcal{O}[N]$). However, since all particles are spread out over the same distance (number of cells), they lack adaptability. If the (constant) cloud size around each particle is large, the resulting grid assignment will overly smooth gradients and lose locality. Conversely, if the cloud is too small, the results around low-density regions will be subject to shot noise and, potentially, the assignment will leave holes in the gridded domain. Thus, an adaptive scheme should be the option to get an appropriate representation of the physical variables on the grid from particle fields.

A common approach in the literature \citep{price2007splash} when dealing with this problem is to spread particle quantities with an smoothed-particle hydrodynamics (SPH) kernel, using an adaptive \textit{smoothing length}, which is different for each particle ($h_i$) and represents the distance to its $k$-nearest neighbor, being $k$ a configurable parameter. In this regard, $k$-d tree's capabilities can be leveraged, significantly accelerating the $k$NN search by using such a query on the tree, which would allow a fast calculation of all $\{h_i \}_{i=1}^N$ distances and identification of the neighbors. As an example, \citet{valles2024vortex} take advantage of a similar technique to accelerate the particle-to-grid assignment of the velocity field, optimizing their Helmholtz-Hodge and Reynolds decomposition algorithm.

\begin{figure*}[h!]
\centering 
\includegraphics[width=1\linewidth]{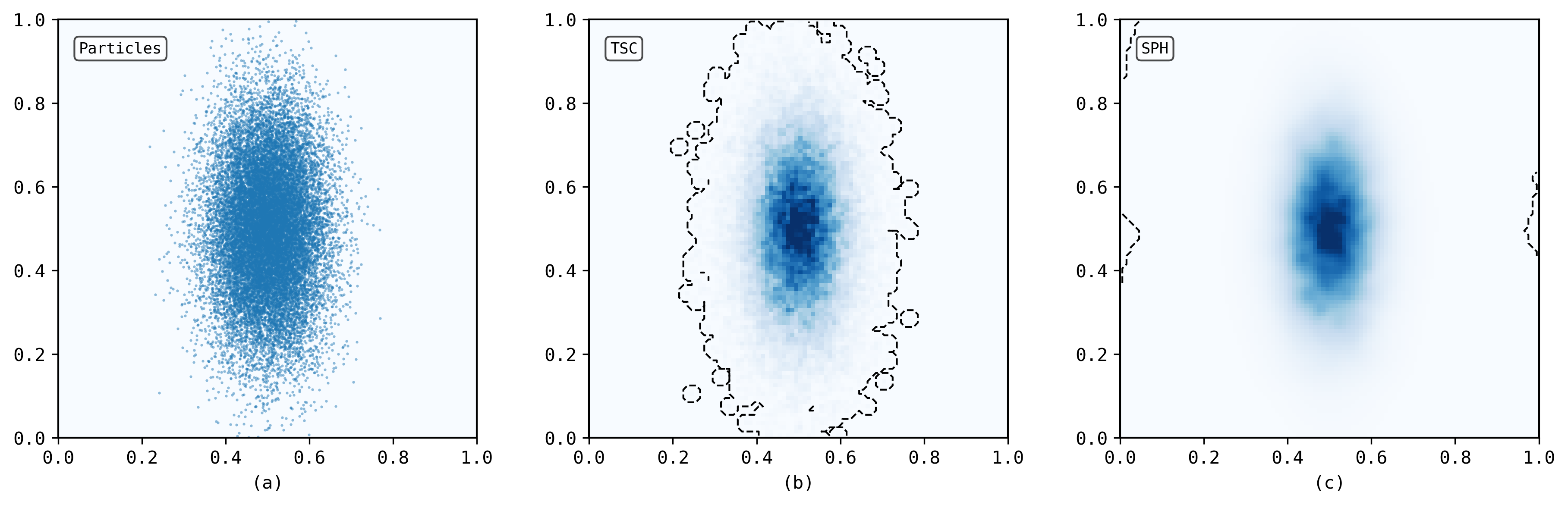}
\caption{2D projection of a distribution of $N = 20000$ particles (a) interpolated onto a $100^3$ grid using the TSC scheme (b) and the adaptive SPH technique (c) accelerated by the $k$-d tree. The dashed contour line separates those cells that have been filled by the particle assignment from those left without any value. The time taken for each method to carry out the particle-to-grid assignment is shown in Table \ref{tab:part2grid}.}
\label{fig:part2grid}
\end{figure*}

To illustrate the process of passing from a particle representation to a grid one, in Fig. \ref{fig:part2grid} we plot a distribution of particles with the corresponding TSC and adaptive-smoothing interpolated mesh fields. As it can be seen, the adaptive option in panel (c) displays a much smoother representation of the number density than the TSC one, being both, by construction, conservative, continuous and differentiable. Besides, the TSC scheme leaves approximately $50\%$ of the box without any value, due to its limited range. This is not a problem for the adaptive interpolation, as it is able to assign a value for most cells\footnote{Note, however, how a full mesh filling is not ensured by this technique. \citet{valles2024vortex} describe a similar methodology that avoids this issue.}. In this sense, apart from providing more visually coherent  depictions of the physical quantities, the \textit{adaptive} approach also yields more reliable and robust representations of these variables on a grid, necessary to carry out calculations such as the ones described before. The only handicap is the fact that each particle smoothing length ($h_i$) has to be calculated. 

In principle, this issue increases the computational cost of performing the particle smoothing. However, as we previously explained, a $k$-d tree can dramatically accelerate the calculations to obtain these lengths by taking advantage of the $k$NN query. So as to demonstrate this fact, we provide Table \ref{tab:part2grid}, displaying the time taken for each method (TSC and SPH using the brute-force and $k$-d tree approaches) to perform the particle-to-grid assignment for different values of $N$. Clearly, though SPH approaches yield better results, using a brute-force approach becomes unmanageable for large $N$. Nevertheless, when using a $k$-d tree to accelerate the processes, computational costs drop considerably. Lastly, it can be seen that, even when using a tree, the adaptive (SPH) method is considerably slower than using a fixed cloud to spread particles (TSC). However, in many applications such as those described above, a more accurate approach is needed, at the expense of spending more resources.

\begin{table}[h!]
\centering
\caption{Time to perform the particle-to-grid interpolation for different methods and number of input points ($N$). Particles are assigned onto a $N_\text{cell} = 100^3$ uniform grid.}
\label{tab:part2grid}
\vspace{0.2cm}
\begin{tabular}{c||c|c|c|}
$N$ & $t_\text{TSC}\, (s)$ & $t_\text{SPH,brute}\, (s)$ & $t_\text{SPH,tree}\, (s)$ \\
 \hline
 10.000 & 0.012 & 6.07 & 0.38 \\
 20.000 & 0.019 & 24.45 & 0.48 \\
 40.000 & 0.030 & 101.80 & 0.61 \\
 80.000 & 0.045 & 428.61 & 0.86 \\
\end{tabular}
\end{table}

\section{Summary}
\label{s:summary}

A new implementation of the $k$-d tree algorithm for computationally intensive applications, \cosmokdtree{}, is presented. The code is fully written in \texttt{Fortran} and the tree construction is parallelized using \texttt{OpenMP} directives. \texttt{Python} bindings are also provided leveraging \texttt{F2PY}. Below, we provide a list of its main features regarding implementation choices:

\begin{itemize}
    \item The sliding-midpoint splitting method is used to build the tree top-down.
    \item A configurable leaf size is used, defaulting to \mbox{$N_\text{leaf} = 16$}. A comprehensive study on choosing the best value for this parameter is provided in \ref{s:appendix.1}.
    \item Tree construction is parallelized leveraging \texttt{OpenMP} \texttt{task} constructs, up to a maximum tree level ($\ell_\text{max}$) is reached and all available threads ($N_\text{CPU}$) are used.
    \item The \textit{Max Heap} and \textit{Min Heap} structures together with efficient bounding box pruning are utilized to accelerate tree traversal during both nearest neighbors and ball queries. 
    \item In case of needing sorted results, a \textit{Quicksort} implementation is utilized.
    \item The algorithm is fully customizable and versatile, as the user can specify the integer size, floating point arithmetics precision, periodic boundary conditions, or dimensionality of the input space at compilation time.
    \item Python bindings are also provided, widening \cosmokdtree{}'s potential use-cases.
\end{itemize}

When all these ingredients are joined together, the resulting $k$-d tree algorithm is fast, both in terms of tree construction and queries, and flexible, allowing it to be accommodated to a broad range of scenarios. We have tested the method against a wide range of scenarios: number of input points, dimensionality, spatial distribution of points and CPU architecture. Furthermore, a comprehensive comparison with other $k$-d tree implementations is also presented:

\begin{itemize}
    \item \cosmokdtree{} achieves a factor of $10-20$ shorter building times than other CPU implementations (\scipy{} \citep{virtanen2020scipy} and \coretran{} \citep{coretran}), displaying similar performance to efficient GPU implementations (\citep{wald2022gpu}  and \torchkdtree{} \citep{torch_kdtree}) executed on high-end consumer graphics cards such as the NVIDIA RTX 3090Ti or RTX 5090.

    \item Across all query benchmarks, \cosmokdtree{} attains the lowest times in general, consistently outperforming other implementations for $k \in (10, 1000)$ and $R \in (10^{-3}, 10^{-1}) L$ and being equally efficient in some extreme cases.

    \item  Such an increase in performance is, to some extent, due to a tradeoff between memory usage and query speed. For instance, \cosmokdtree{}'s implementation uses twice as much memory as \scipy{}'s if double-precision floats are considered. However, in this way, memory-locality and thread-safety are ensured, with a resulting tree structure that is completely self-contained and, hence, easy to embed in wider codes.
\end{itemize}

From these points, we conclude that \cosmokdtree{} stands as one of the fastest $k$-d tree implementations on CPU while matching or marginally falling behind GPU implementations run on high-end consumer hardware. Though there is room for more advanced GPUs to further outperform our implementation, we do not pursue such exploration in this manuscript, as the design purpose of the library is not to be the fastest $k$-d tree, but a compromise between flexibility and efficiency. To exemplify \cosmokdtree{}'s feasible extensive use cases, two applications are displayed, highlighting its ability to accelerate analysis tools for processing vast datasets: 
\begin{enumerate}
    \item In the first, a friends-of-friends (clustering) algorithm is implemented in \texttt{Fortran} and run with three different configurations to search for close neighbors: one uses brute-force, another leverages a uniform auxiliary grid to limit neighbor searches to nearby cells, and the third employs our $k$-d tree implementation. While both the grid and tree techniques considerably outperform the brute-force method, for large datasets ($N> 10^6$) the tree search vastly surpasses the other approaches, yielding an optimal $\mathcal{O}(N \log N)$ time complexity.

    \item Regarding the second, the particle smoothing techniques, commonly used to interpolate particle fields to a grid, are reviewed. One of the most refined methods employs an SPH kernel using the distance to the $k$-nearest neighbor (being $k$ configurable) as smoothing length. For very large numbers of particles, calculating all particle smoothing lengths can be too expensive. However, the $k$-d tree $k$NN query can be harnessed to efficiently obtain these distances, considerably reducing the computational cost of performing the particle-to-grid interpolation of variables. In this sense, several analysis tools (e.g., see \citet{valles2024vortex, monllor2025avism}) take advantage of a $k$-d tree to rapidly obtain a continuous (grid) representation of physical quantities, necessary to perform their calculations.
\end{enumerate}

\cosmokdtree{}'s ultimate goal is not to achieve the best performance (which often comes at the expense of reducing usability), but a compromise between maximizing efficiency and flexibility. In this direction, we believe our objectives have been fulfilled, as proven by previous points describing the benchmark results and applicability. Within this context, given \cosmokdtree{}'s features and the broad range of astrophysical scenarios in which it could be applied, we consider it a convenient and optimal tool to accelerate computationally intensive tasks in astrophysics, being them in the form of analysis and post-processing codes for vast datasets (modern simulation outputs, large object catalogues, etc.), data-generating algorithms (such as simulation codes) or any computational application needing fast spatial queries (e.g., ray-tracing). We believe it could be particularly interesting for already existing tools needing further optimization to handle large data volumes. We publicly release the code and tests in the corresponding GitHub repository\footnote{\url{https://github.com/oscarmonllor99/cosmokdtree}}.

\section*{Acknowledgements}

 We thank the anonymous referee for the comments that helped improving the quality of this manuscript. This work has been supported by the Agencia Estatal de Investigación Española (AEI; grant PID2022-138855NB-C33) and by the Generalitat Valenciana (grant PROMETEO CIPROM/2022/49).
 OM and DV acknowledge support from Universitat de València through Atracció de Talent fellowships. DV acknowledges additional support from the ERC CoG $\vec{B}$ELOVED, GA n. 101169773.

\bibliographystyle{elsarticle-num-names} 
\bibliography{kdtree.bib}

\begin{appendix}
\numberwithin{figure}{section}

\section{Leaf size sweet spot}
\label{s:appendix.1}

\begin{figure*}[h!]
\centering 
\includegraphics[width=1\linewidth]{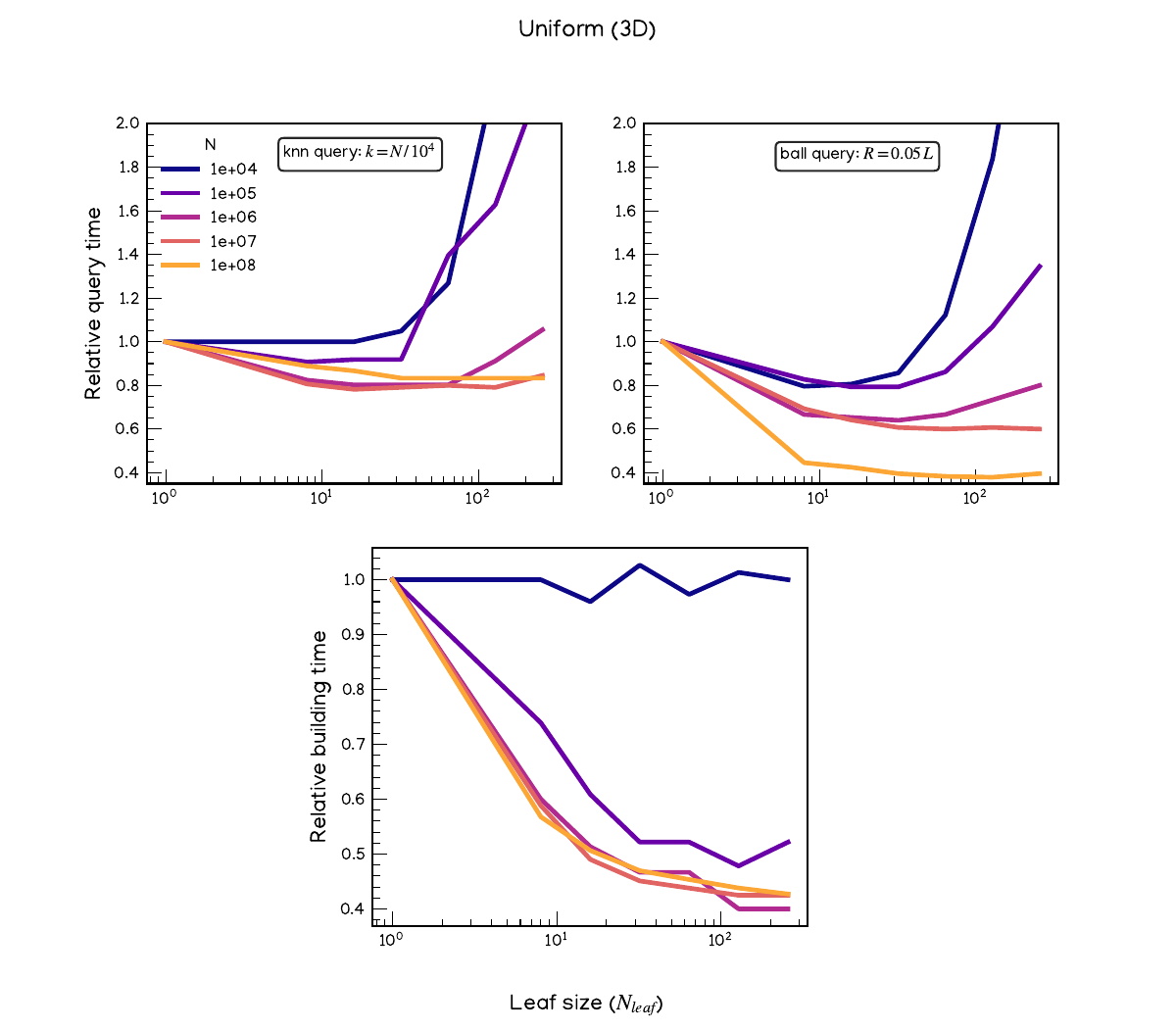}
\caption{Relative query time (with respect to the $N_\text{leaf} = 1$ case) against leaf size for the $k$-nearest neighbors search (\textit{upper left}) and ball search (\textit{upper right}) for different sizes ($N$) of input data. In the \textit{lower panel}, we display the relative tree construction time (relative to $N_\text{leaf} = 1$) against leaf size, also for different numbers of input points.}
\label{fig:leafsize_test_3D}
\end{figure*}

\begin{figure*}[h!]
\centering 
\includegraphics[width=1\linewidth]{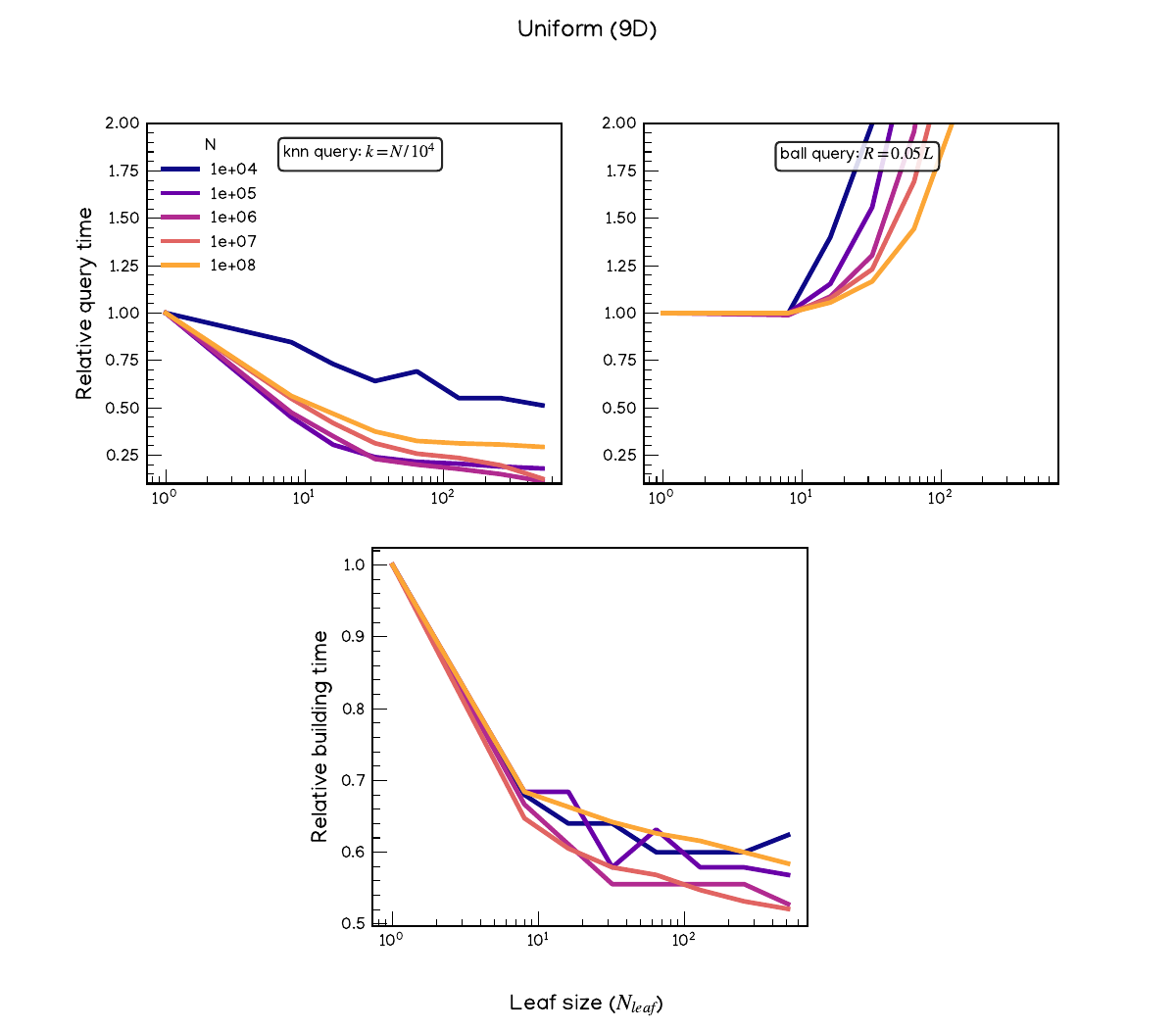}
\caption{Same as Figure. \ref{fig:leafsize_test_3D}, but for a 9-dimensional uniform distribution of points.}
\label{fig:leafsize_test_9D}
\end{figure*}

In order to find the leaf size ($N_\text{leaf}$) that makes our $k$-d tree perform the best in terms of tree construction and query times, we have run two test from which we can precisely extract which is the code performance scaling for varying leaf size ($N_\text{leaf}$) depending on the number of input points ($N$) and dimensionality ($D$). 

In the first, we vary $N$ from $10^4$ to $10^8$ in powers of ten, distributed uniformly in a 3-dimensional cubical box of side length $L$. At the same time, we modify our $k$-d tree's leaf size from unity to $8,16,32,\dots$ on so on, in powers of two. For all combinations of $N$ and $N_\text{leaf}$, we measure the tree construction time, the ball query time for a radius $R = 0.05 \, L$ and the $k$-nearest neighbors search time for $k = N/10^4$. We set $k$ to a fixed fraction of the input dataset size to ensure the leaf size effect can be noticeable at all $N$.

In Figure \ref{fig:leafsize_test_3D}, the relative query and construction times, with respect to the $N_\text{leaf} = 1$ case, are displayed for different values of leaf size and number of input points for the tree. As expected, tree construction time is reduced for increasing $N_\text{leaf}$ (except for $N = 10^4$, where it is already fast). Regarding queries, increasing the leaf sizes can help to accelerate them significantly (more than two times faster in some cases). Nevertheless, and especially for small datasets, this can quickly become counterproductive, as queries can become exponentially slower, since brute force starts taking over. The main reason for this is the fact that smaller datasets will produce shallower trees in which the effect of the leaf size can help little to accelerate the process of traversing the tree to find the neighbors of a point. In general, if the leaf size has a value of $N_\text{leaf}$ and the number of input points is $N$, the total tree level can be approximately known before construction:
\begin{equation}
    \ell_\text{tot} \approx \log_2  \bigg(\frac{N}{N_\text{leaf}} \bigg),
\end{equation}
with $\ell$ denoting the maximum tree level. 

Regarding the second test, we repeat the same configuration as before, but this time in a 9-dimensional space. Results are shown in Figure \ref{fig:leafsize_test_3D}. In this case, the $k$NN query is always accelerated (for all $N$) for increasing $N_\text{leaf}$, while the ball search quickly becomes inefficient when $N_\text{leaf}$ surpasses the $10-20$ range (depending on $N$). Thus, while in the 3D case, the ball query is the one getting the most benefits from increasing $N_\text{leaf}$, the opposite happens for higher dimensions (9D), where $k$NN query times monotonically decrease with increasing $N_\text{leaf}$. Such dissimilar results highlight the complexity of choosing a proper leaf size, depending on $N$ and $D$. 

Since, independently of $N$ and $D$, tree construction times are significantly reduced for $N_\text{leaf} = 10-20$ and $k$NN and ball queries always get more efficient in this range, we consider $N_\text{leaf} = 16$ the optimal value for this parameter and, hence, we set it as \cosmokdtree{}'s default. 

\section{$k$-d tree construction algorithms}
\label{s:appendix.2}
Construction algorithms can be divided into two main classes: top-down or bottom-up. The first builds the tree from the whole original dataset to the leaves (top-down), while the second builds the tree from the child nodes up to the original dataset (bottom-up). All construction methodologies must produce a $k$-d tree structure which is balanced (tree depth is limited) and complete (Eq. \eqref{eq:left_right} is always fulfilled).

\cosmokdtree{}, \scipy{}, \coretran{}, \citep{brown2014building}, \cite{wehr2018parallel} and \citep{wald2022gpu} $k$-d tree constructions are among many examples belonging to the top-down class. They build the tree recursively from the original dataset until all branches find a leaf (see our detailed description above). The logic behind the algorithm is simple and, thus, the effort is put on efficiently performing the partitioning at each level. In principle, the main caveat incurs parallelization scaling, as level-based concurrency results in the first dataset ($S_0$) being divided only by one thread, the two child nodes by two thread, the four grandchildren by four thread, etc. Taking into account that the first splits are the most expensive (the ones with more points to process), parallelization scaling which such an approach is halted by the inherent sequential structure of the algorithm at the top levels. However, the parallelization effort can be redirected to the partitioning step, which, depending on the underlying algorithm, can yield better scaling than the level-based approach, though the overall procedure gets significantly more complex. The last three references provided above follow this line, being able to efficiently parallelize top-down tree constructions in GPUs.

On the other hand, Ref. \cite{karras2012maximizing} would be an example from the bottom-up technique. This methodology was devised to avoid the sequential bottlenecks described for the top-down approach. The tenet behind it is to build all nodes at once (thus eliminating recursion) and afterwards find the correct position of each node inside the structure so that the resulting tree is balanced and complete. The steps that have to be carried out to make this inherently sequential problem suited for parallelization are substantially more complex than the usual top-down partitioning, adding overhead to the overall tree construction. However, the extra calculations are offset by the improvement in parallelization, which can vastly accelerate the node processing part.

\end{appendix}

\end{document}